\documentclass[useAMS,usenatbib]{mnras}
\usepackage{times,graphics,graphicx,multicol,amssymb} 

\def\aap{A\&A}

\def\apj{ApJ}

\def\apjl{ApJ}
\def\mnras{MNRAS}
\def\araa{ARA\&A}
\def\aj{AJ}

\def\apjs{ApJS}

\def\ssr{Space Sci. Rev. }

\def\lesssim{\mathrel{\hbox{\rlap{\hbox{\lower4pt\hbox{$\sim$}}}\hbox{$<$}}}}
\def\gesssim{\mathrel{\hbox{\rlap{\hbox{\lower4pt\hbox{$\sim$}}}\hbox{$>$}}}}

\def\vect#1{{\mathbfit{#1}}}

\begin{document} 

\author[A. Morandi and M. Sun]
{Andrea Morandi${}^1$\thanks{E-mail: andrea.morandi@uah.edu}, Ming Sun${}^1$\\
$^{1}$ Physics Department, University of Alabama in Huntsville, Huntsville, AL 35899, USA\\
}

\date{}

\title[Probing dark energy via cluster outskirts] %45 characters including spaces
{Probing dark energy via galaxy cluster outskirts}
\maketitle
 
\begin{abstract}
We present a Bayesian approach to combine {\em Planck} data and the X-ray physical properties of the intracluster medium in the virialization region of a sample of 320 galaxy clusters ($0.056<z<1.24$, $kT\gesssim 3$ keV) observed with {\em Chandra}. We exploited the high-level of similarity of the emission measure in the cluster outskirts as cosmology proxy. The cosmological parameters are thus constrained assuming that the emission measure profiles at different redshift are weakly self-similar, that is their shape is universal, explicitly allowing for temperature and redshift dependence of the gas fraction. This cosmological test, in combination with {\em Planck}+SNIa data, allows us to put a tight constraint on the dark energy models. For a constant-$w$ model, we have $w=-1.010\pm0.030$ and $\Omega_m=0.311\pm0.014$, while for a time-evolving equation of state of dark energy $w(z)$ we have $\Omega_m=0.308\pm 0.017$, $w_0=-0.993\pm0.046$ and $w_a=-0.123\pm0.400$. Constraints on the cosmology are further improved by adding priors on the gas fraction evolution from hydrodynamic simulations. Current data favour the cosmological constant with $w\equiv-1$, with no evidence for dynamic dark energy. We checked that our method is robust towards different sources of systematics, including background modelling, outlier measurements, selection effects, inhomogeneities of the gas distribution and cosmic filaments. We also provided for the first time constraints on which definition of cluster boundary radius is more tenable, namely based on a fixed overdensity with respect to the critical density of the Universe. This novel cosmological test has the capacity to provide a generational leap forward in our understanding of the equation of state of dark energy.
\end{abstract}

\begin{keywords} 
methods: data analysis -- methods: statistical -- galaxies: clusters: general -- cosmology: observations -- large-scale structure of Universe -- X-rays: galaxies: clusters.
\end{keywords}

\section{Introduction}\label{intro}

The hierarchical large-scale structure scenario (LSS) provides a picture where gravity is constantly pulling lumps of matter together to form increasingly larger structures, with galaxy clusters sitting atop this hierarchy, i.e. at the intersection of filaments. Clusters, thanks to their position at the high end of the cosmic mass power spectrum, are sensitive proxies of cosmic evolution and a unique probe of mass distribution on large scales. 

In particular, measurements of cluster growth via their mass function \citep{vikhlinin2009c} and relative baryon budget \citep{mantz2014} are highly sensitive tests of the geometry and matter content of the Universe. These constraints are complementary to measurements of the expansion history from Type Ia supernovae (SNIa), the cosmic microwave background (CMB) temperature and polarization spectra, and baryon acoustic oscillations (BAO). The combination of observables referring to the expansion history and geometry of the Universe (e.g. baryon fraction in galaxy cluster, SNe, BAO), and those referring to structure growth (e.g. the evolution of the cluster abundance) can be therefore used to constrain the cosmology, since a consistency relation between all these observables must occur in any theory which aims at explaining the cosmic acceleration of the Universe.

Recently, the improvements of the data quality of the observations, with the accumulation of observational data from CMB measurements, LSS surveys, and SNIa observations, have allowed one to constrain the cosmological parameters with a greatly improved accuracy, at few percent level. In particular, the Planck Collaboration has released the first cosmological papers providing the highest resolution, full sky maps of the CMB temperature anisotropies. These results suggest that a model in which the dark energy is a cosmological constant ($w =-1$) is preferred \citep{planck2015a}. These findings have been independently corroborated by SNIa \citep{betoule2014} and galaxy cluster growth analyses \citep{vikhlinin2009c}. However, observational signatures of such deviations of $w$ from $-1$ are very small, and hence the measurements are prone to statistical and systematic errors. Therefore, current cosmological constraints are still limited by the quality of the data and systematics in complementary proxies of the expansion history of the Universe. 

For example, the use of clusters as accurate cosmological probes requires to unearth and disentangle the cosmological and astrophysical signatures since, in practice, the cluster observables are customarily locked to the baryonic component. With respect to the dark matter (DM), cluster baryons are indeed subject to more complex (and currently still uncertain) physical processes (e.g. galaxy formation and feedback) besides the gravity. This can clearly limit the use of clusters as cosmological proxies. While it is paramount to improve our modelling of clusters, it is also crucially important that the dark energy constraints at this level of accuracy are obtained from combination of several independent techniques and data sets. This not only reduces systematics but also improves statistical accuracy by breaking degeneracies in the cosmological parameter constraints.

One cosmological test via galaxy clusters which has been previously proposed \citep[e.g.][]{arnaud2002} hinges on the self-similarity. Indeed, for an assumed cosmology, similarity of shape in cluster radial quantities has been recently shown in X-ray observations \citep{arnaud2002,arnaud2010,eckert2012,morandi2015}, with weak-lensing \citep[][]{postman2012} and SZ \citep{plagge2010,planck2013}. These homogeneous physical properties of the intracluster medium (ICM) are theoretically expected since the scale-free DM collapse drives the evolution of halo concentration and the physical properties of the baryons across cosmic time \citep{limousin2013}. The ensuing similarity yields both a universal DM distribution and simple global scaling relations between the observables \citep{kaiser1986,voit2005a,arnaud2010}, retaining the fingerprints of the underlying cosmology, gravity but also astrophysics that govern their formation and evolution. Given the current large samples of clusters e.g. from {\em Planck} or {\em Chandra}, the most striking outgrowth of self-similarity is the ability to greatly exploit this large amount of data, by stacking the X-ray/SZ signal to study the virialization region of clusters \citep{plagge2010,eckert2012,planck2013,morandi2015}.

In this respect, in the present paper we investigated cosmological constraints on the expansion history and geometry of the Universe to study dark energy models, by exploiting the high-level of similarity of X-ray emission measure ($EM$) profiles in the outskirts of the hot cluster population. In a companion paper \citep{morandi2015}, we stacked the emission measure profiles $EM\propto \int n_e^2\, dl$ of a X-ray {\em Chandra} sample of 320 clusters to detect a signal out to and beyond $R_{200}$. In that work we proved that the physical properties of clusters are indeed universal outside of the core ($R \gesssim 0.2 R_{200}$) once scaled according to a certain radius e.g. $R_{200}$, where $R_{200}$ is the radius within which the mean total density is 200 times the critical density of the Universe. The idea of this cosmological test is that the X-ray $EM$ profiles, once rescaled by the self-similar predictions, can be used as `standard candle' to constrain the cosmological parameters, being a sensitive proxy of the underlying cosmology via both the angular distance and the critical density of the Universe $\rho_{c,z}$. The correct cosmology is thus constrained assuming weak self-similarity of $EM$ for the outskirts of hot clusters, that is the various $EM$ profiles at different redshifts are homologous, explicitly allowing for temperature and redshift dependence of the gas fraction $f_{\rm gas}$. In fact, at $R>R_{1000}$, even groups and poor clusters become more self-similar \citep[e.g.][]{sun2012}. Using the wealth of data in hand, both independently of and in combination with complementary data sets, we present constraints on the dark energy models within the $\Lambda$ cold dark matter ($\Lambda$CDM), in particular to determine whether there is evidence for dynamical dark energy with equation of state (EoS) parameter $w\neq -1$.

The paper is organized as follows. In Section \ref{swefswfelfsim} we describe the self-similar model, cosmological framework and we describe the origin of the constraints on the cosmological parameters from the observables. In Section \ref{swe222sim} we present our data set, while in Section \ref{rjmcmc1} we present the statistical method used to infer both model and physical parameters. In Section \ref{swefffsim} we present results on the EoS of dark energy, a model comparison (Section \ref{sweg6fsim}), the analysis of the systematics (Section \ref{sgnegneg53}) and the potential for improvements in
cosmological constraints from future surveys (Section \ref{swg36fwf2z1ded}). Throughout this work we assume the flat CDM model, with matter density parameter $\Omega_{\rm m}$, EoS parameter $w$, and Hubble constant $H_{0}=100h \,{\rm km\; s^{-1}\; Mpc^{-1}}$. The factor $E_z=H(z)/H_0$ is the ratio of the Hubble constant at redshift $z$ to its present-day value. Unless otherwise stated, we report the errors at the 68.3\% confidence level.

\section{Cosmology via outskirts of galaxy clusters: methodology}\label{swefswfelfsim}

In this section we will review the basic idea of the self-similar model for cluster outskirts (Section \ref{selfsim}), emphasizing how clusters are identical objects with respect to a critical overdensity and when scaled by their temperature. We then outline our current cosmological framework, in particular with respect to the dependence of the cluster observables on the underlying cosmology (Section \ref{cosmddd2}). Next, we discuss how to pin down the cosmological parameters through the observables (Section \ref{cosmwdded2}), comparing our constraints to those from other independent measurements (e.g. CMB, LSS, SNIa). Finally, we present a generalized self-similar model which accounts for the impact of the baryon component at the top of the evolution of the cosmic expansion history (Section \ref{coerg5uded2}).

\subsection{The self-similar model}\label{selfsim}

In the current hierarchical structure formation model, galaxy clusters arise from the gravitational collapse of rare high peaks of primordial density perturbations, when their linear evolution reaches the collapse threshold $\delta_c = 1.686$. On large scales, the majority of the baryonic component is in the form of X-ray emitting ICM and is expected to follow the distribution of the gravitationally dominant DM. The self-similar model \citep[see, e.g.,][]{kaiser1986} predicts that cluster gas profiles for a given mass (or peak height) appear universal when they are scaled according to a certain radius (e.g. $R_{200}$) with respect to the reference background density of the Universe \citep[see, e.g.,][]{voit2005a}. The underlying idea is that gravity is the only responsible for the observed values of the physical properties of clusters and the ICM is heated only by the shocks associated with the collapse. Hence, clusters are identical objects when scaled by their mass or temperature and power-law scaling relations are expected between the observables. Baryonic physics, such as radiative cooling, star formation and feedback might break self-similarity, but their effects are mostly confined within the cluster core.

Assuming the spherical collapse model for the DM halo and the equation of hydrostatic equilibrium to describe the distribution of baryons into the DM potential well, in the self-similar model the cluster mass at an overdensity $\Delta$ (e.g. $\Delta=500,200$) and temperature are related by $E_{z} M_{\rm tot} \propto T^{3/2}$. The overdensity $\Delta$ is calculated with respect to the reference background density of the Universe, customarily the critical (or mean) density (see Section \ref{swcdfffrm3ed} for further discussion). So we have $ R_{\Delta} \propto {\left({M/(\rho_{c,z}\Delta)}\right)}^{1/3} \propto T^{1/2}E_{z}^{-1}\Delta^{-1/2}$. From the previous equation it follows that:
\begin{equation}
EM(x) \propto \int n_e^2\; dl \propto M_{\rm gas}^2/R_{\Delta}^5 \propto  f_{\rm gas}^2 \; E_z^3 T_X^{1/2}
\label{dweded}
\end{equation}
with $x=R/R_{\Delta}$, $f_{\rm gas}=M_{\rm gas} / M_{\rm tot}$ being the gas mass fraction.

In this self-similar model, galaxy clusters of all masses and redshift are identical objects when scaled according to equation (\ref{dweded}). This is referred as {\it strong self-similarity} \citep[][]{kaiser1986}, and sets the powerlaw slopes of the scaling relations, which are not predicted to evolve with redshift and are independent of the cluster mass.

As can be seen from equation (\ref{dweded}), clusters of given temperature appear smaller ($R_{\Delta}\propto E_z^{-1}$, with e.g. $\Delta_z=\Delta=200$) and brighter ($EM(x=0) \propto E_z^3$) with increasing redshift, following their self-similar evolution. This translates into a surface brightness dimming $S_x\propto E_z^3\, (1+z)^{-4}$, wich corresponds to an only modest attenuation at $z\sim0.6-1$ (a factor of $\sim2$-3) with respect to the cosmological dimming factor $(1+z)^{4}\sim6-16$. 

\subsection{Cosmological framework}\label{cosmddd2}
We refer to $\Omega_{\rm m}$ as the {\it total matter} density and to $\Omega_{\Lambda}$ as the {\it dark energy} density. Therefore, we adopt the Einstein equation in the form $\Omega_{m} + \Omega_{\Lambda} +\Omega_k = 1$, where $\Omega_k$ accounts for the spatial curvature of the Universe. 

We remember that in an hierarchical scenario of structure formation cluster outskirts are thought to be self-similar when they are normalized to an overdensity $\Delta$ with respect to the critical density, $\rho_{\rm c,z} = 3 H_0^2E_z^2/ (8 \pi G)$. Thus, from equation (\ref{dweded}) the (scaled) $EM$ profiles depend on the assumed cosmological parameters via the angular distance $D_{\rm a}$ used to convert angular radii $\theta$ to physical radii, and via the factor $E_z=H(z)/H_0$ appearing both in the normalization of the profiles and in $R_{\Delta}$ (i.e. $R_{\Delta}\propto E_z^{-1}$). 

From, e.g., \cite[][cf. equation 25]{carroll1992}, we can then write the angular diameter distance and the factor $E_z$ as:
\begin{eqnarray}
% D_{\rm a} & = &\frac{c}{H_0 (1+z)}  \int^z_0 \frac{d \zeta}{E_z(\zeta)},\nonumber \\
D_{\rm a} & = &\frac{c}{H_0 (1+z)} \frac{S(\omega)}{|\Omega_k|^{1/2}},   \nonumber  \\
\omega &= & |\Omega_k|^{1/2} \int^z_0 \frac{d \zeta}{E(\zeta)} \nonumber  \\
E_z & = & \left[\Omega_{\rm m} (1+z)^3 + \Omega_{ k}(1+z)^2 + \Omega_{\Lambda} \lambda_z \right]^{1/2} \nonumber \\
\lambda_z & = & \exp\left(3 \int_0^z \frac{1+w(z)}{1+z} dz\right)\,,
\label{eq:ez}
\end{eqnarray}
where $S(\omega)$ is sinh$(\omega)$, $\omega$, $\sin(\omega)$ for $\Omega_{k}$ greater than, equal to and less than 0, respectively. Note that in our analysis we neglect the energy associated with the cosmic radiation, $\Omega_r \simeq 4.16 \times 10^{-5}(T_{\rm CMB}/2.726\,K)^4$.

The variation with redshift of $D_{\rm a}$ and $E_z$ depends on $\Omega_m$, $\Omega_{\Lambda}$, $\Omega_k$ and $w(z)$. Therefore, assuming the self-similar evolution model and similarity (that is universal shape) of $EM$ at various redshifts (in the correct cosmology), we can infer the cosmological parameters. We refer to Section \ref{cosmwdded2} for a more detailed discussion on the dependence of the (scaled) $EM$ on the cosmological parameters.

The properties of dark energy are commonly characterized by its equation-of-state parameter $w=P/\rho$, defined as the ratio between the pressure and the energy density in the EoS of the dark energy component \citep{caldwell1998,wang1998}. In our baseline model we assume that the dark energy is a cosmological constant with current density parameter $\Omega_{\Lambda}=1-\Omega_m$, and assume that the dark energy does not interact with other constituents other than through gravity. When considering a dynamical dark energy component, we parametrize the EoS either as a constant $w$ or as a function of the redshift. Hereafter we shall consider parametrizations of $w$ encompassing a pressure-to-density ratio $w$ constant and time-evolving in a flat Universe ($\Omega_k = 0$). In particular, we will consider the case of a cosmological constant (standard $\Lambda$CDM scenario), which requires $w=-1$; the model with a constant EoS by taking $w(z)=w_0 $ (model $w$CDM); finally, for dark energy models with a time-evolving $w(z)$ we choose the popular EoS parametrization $w(z)=w_0+w_az/(1+z)$ given by \citet[][]{chevallier2001} (model $w_z$CDM). We shall also discuss the case of a non-flat Universe, i.e. $\Omega_k\neq 0$.

Moreover, the problem is to recover simultaneously both the cosmological model and its relative parameters, and to perform a model selection as well: a convenient approach by means of Bayesian inference has been presented in Section \ref{rjmcmc1}.

\subsection{From observables to cosmology}\label{cosmwdded2}

In this section we discuss how we can pin down the cosmological parameters from cluster observables. 

Our raw observables are represented by the X-ray surface brightness $S_x(\theta)$ (expressed in counts s$^{-1}$ arcmin$^{-2}$, $\theta$ being the angular distance) and global spectroscopic temperatures. These raw quantities can be regarded as {\it cosmological independent}, since they are direct measurements. However, different clusters show significant different ranges of X-ray surface brightness both in normalization and angular extent. For example, a nearby cluster like Coma extends out to a few degrees, while high-$z$ clusters have a typical size of the order of a few arcminutes.

\begin{figure*}
\begin{center}
% \epsscale{.85}
\hbox{
\includegraphics[scale=0.43]{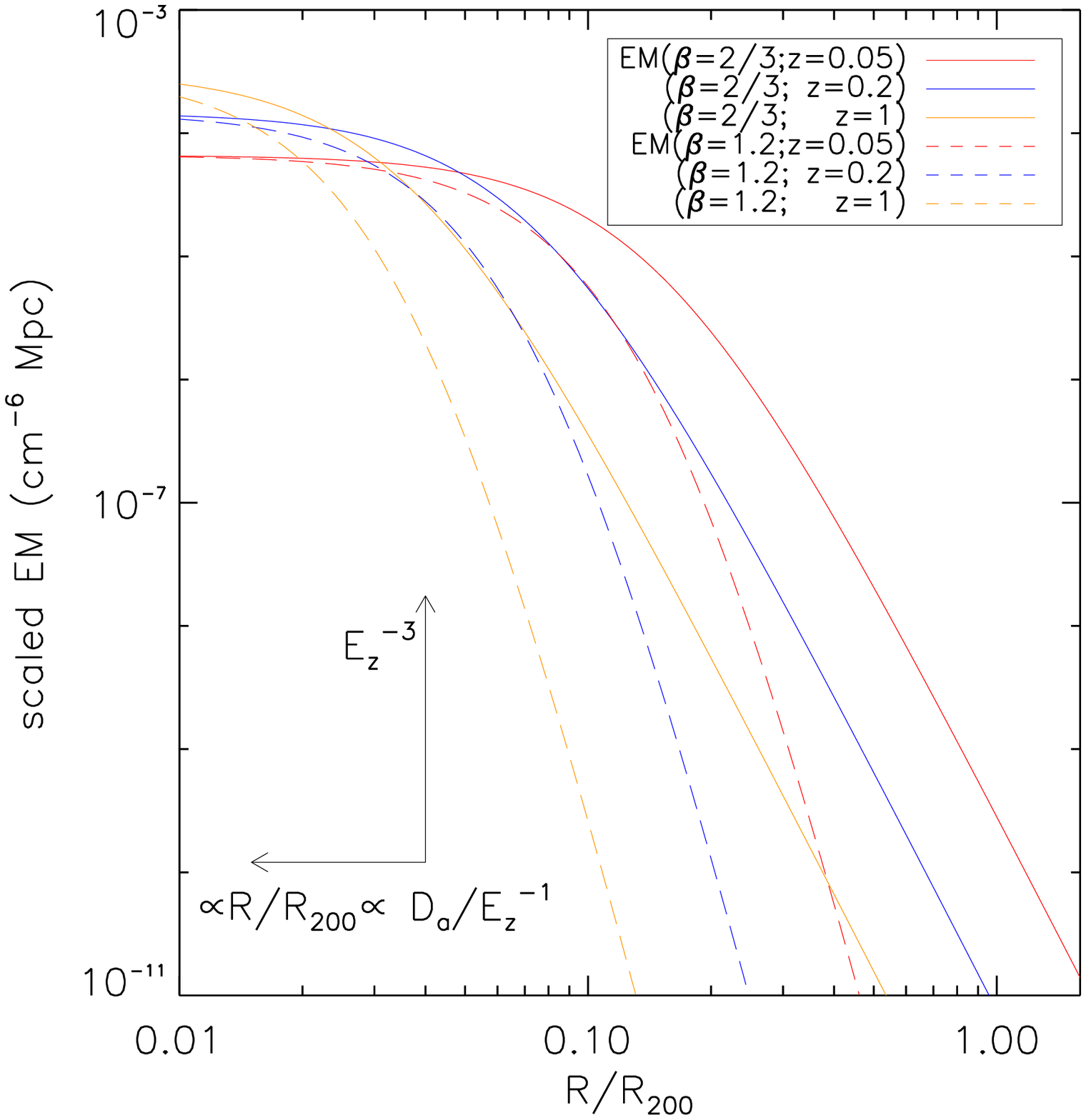}
\includegraphics[scale=0.43]{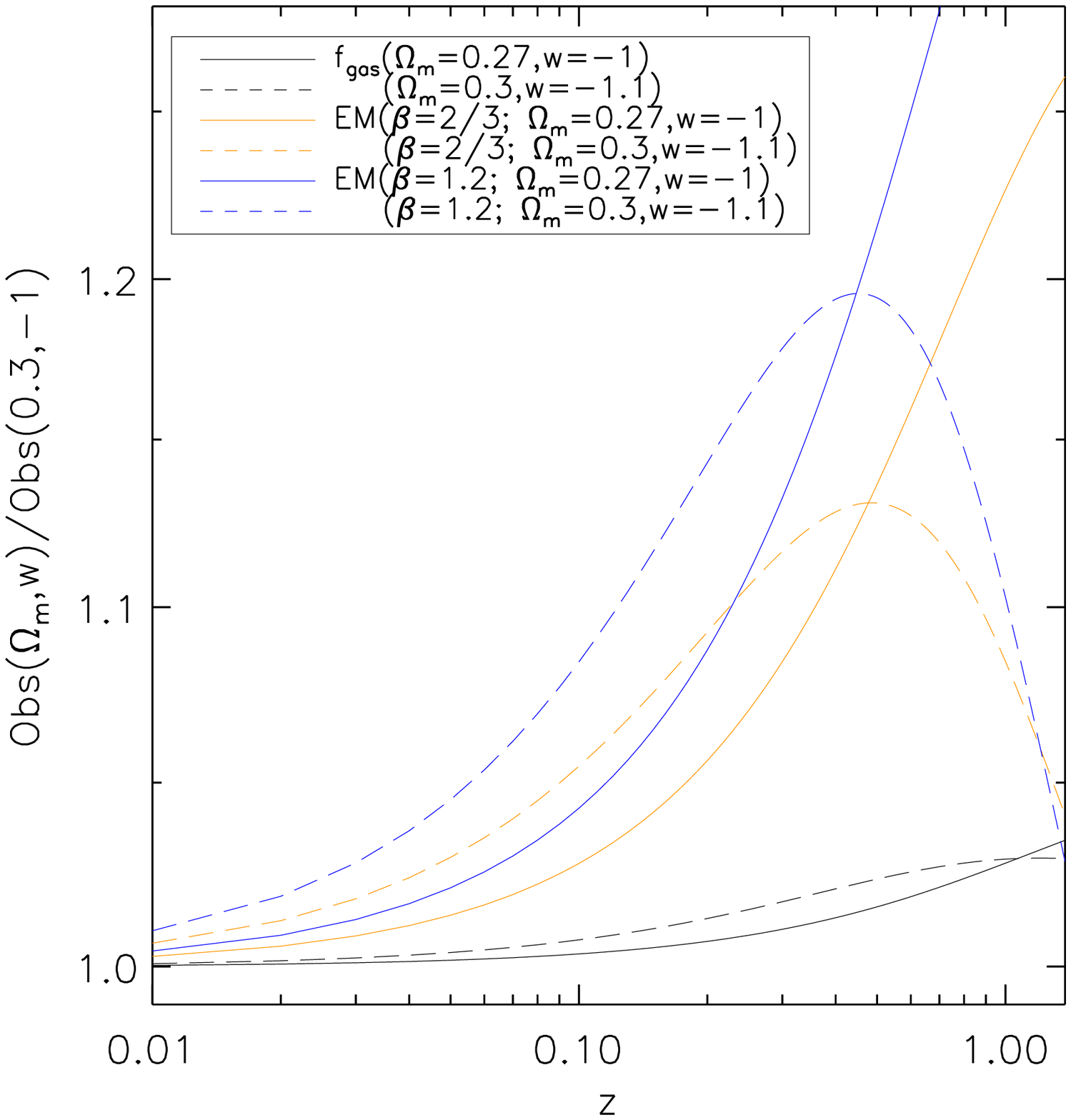}
}
\caption{Left-hand panel: toy model (a $\beta$-model profile with slopes $\beta=2/3$ and 1.2) describing the scaled emission measure $\widetilde{EM}(x)$, i.e. {\em renormalized} by $E_z^{3} f_{\rm gas}^{2} T_X^{1/2}$ and {\em scaled} by $R_{200}$. The $EM$ profiles have been scaled according to the standard self-similar model (equation \ref{dweded2}), thus introducing a dependence upon the cosmology via $E_z=H(z)/H_0$ and the angular diameter distance $D_a$. The $\widetilde{EM}(x)$ from nearby clusters at $z=0.05$ in a {\it reference cosmology ($\Omega_m=0.3,w = -1$)} is compared with the predictions of $\widetilde{EM}(x)$ at $z = 0.2$ and $z=1$ in a {\it different dark energy model ($\Omega_m=0.3,w = -1.1$)}, thus introducing unphysical departures from self-similarity. Changing the underlying cosmology introduces a shift upwards ($\propto E_z^{-3}$) and towards smaller radii ($\propto R/R_{200}\propto D_a / E_z^{-1}$) of the scaled $EM$ profile, with the arrows in the figure showing the magnitude and direction of this shift. By playing the reverse game, i.e. assuming similarity of the $\widetilde{EM}(x)$ profiles at different $z$ (that is the shape of $\widetilde{EM}(x)$ is universal) and for $x\gesssim0.2$, we can recover the {\it correct} cosmology \citep[e.g.][]{arnaud2002}. The magnitude of the scaling and renormalization has been increased by a factor of 2(4) for $z=0.2(z=1)$ for visualization purposes. Right-hand panel: sensitivity to $\Omega_m$ and $w$ of different cosmological proxies $Obs$. We considered the measured gas fraction $f_{\rm gas}=M_{\rm gas} / M_{\rm tot}\propto D_{\rm a}(\Omega_{\rm m}, w)^{3/2}$ and $\widetilde{EM}(\tilde x)$ at a fixed radius $\tilde x$, for $\tilde x\gesssim1$ (equation \ref{dweded2}) as cosmological proxies $Obs$ with respect to a reference cosmology ($\Omega_{\rm m}=0.3,w=-1$).}
\label{T-z}
\end{center}
\end{figure*}

The X-ray surface brightness profiles have been then scaled according to the standard self-similar model (equation \ref{dweded2}), thus introducing a dependence upon the cosmology via the expansion history factor $E_z=H(z)/H_0$ and the angular diameter distance $D_a$ (Fig. \ref{T-z}). Hence the rescaled emission measure $\widetilde{EM}(x)$ (see below for its definition) is {\it self-similar} in the correct cosmology, that is the shape of $\widetilde{EM}(x)$ is universal for clusters with different temperatures and redshifts.

More in details, here we outline the dependence of $\widetilde{EM}(R/R_{200})$ on the cosmological parameters. Without loss of generality, we can assume that the raw $EM$ profile has a powerlaw shape in the external regions, i.e. $EM(x)\propto x^{1-6\beta}$, with $\beta=\beta(x)$ increasing with the cluster-centric radius $x=R/R_{200}$. We define the scaled emission measure as $\widetilde{ EM}(x)=EM(x)E_z^{-3} f_{\rm gas}^{-2} T_X^{-1/2}$ exploiting the self-similar predictions \citep[see, e.g.,][]{kaiser1986,arnaud2002}. We can thus highlight the dependence of $\widetilde{EM}(x)$ on the cosmological parameters:
\begin{equation}
\widetilde {EM}(x)  \propto  T_X^{-1/2} f_{\rm gas}^{-2} E_z^{-3+(1-6\beta)}\; D_{\rm a}^{1-6\beta}
\label{dweded2}
\end{equation}
with $\theta=x\, R_{\Delta}/D_{\rm a}$, where $f_{\rm gas}$ is the gas fraction. The expansion history factor $E_z=H(z)/H_0$ and the angular diameter distance $D_{\rm a}$ are the proxies of the cosmology, depending on $\Omega_m$ and $w=w(z)$.

In Fig. \ref{T-z} (left), we present a toy model showing the dependence of the scaled emission measure $\widetilde{EM}(x)$ (equation \ref{dweded2}) on the cosmological parameters (equations \ref{dweded2} and \ref{eq:ez}). Changing the underlying cosmology produces a shift along the vertical axis ($\propto E_z^{-3}$) and the horizontal axis ($\propto R/R_{200}\propto D_a / E_z^{-1}$) of the scaled $EM$ profile, thus introducing unphysical departures from self-similarity. By playing the reverse game, i.e. assuming similarity of the $\widetilde{EM}(x)$ profiles at different $z$ (that is the shape of $\widetilde{EM}(x)$ is universal) and for $x\gesssim0.2$ (excluding the inner volumes), we can recover the correct cosmology, following the idea successfully applied by \citet{arnaud2002}. 

The scaled $EM$ profiles are not used as simple distance proxy, in the strict sense, depending on both the angular distance, via the conversion of angular radius into physical radius, and also upon the cosmology via the cluster growth term $E_z=H(z)/H_0$, which describes their evolution with $z$ in the self-similar model. Unlike distance-determination methods (e.g. SNIa), galaxy clusters can be thought more like `cosmic buoys', anchored to the evolution of the background cosmology in a very simple way provided that they are self-similar objects (e.g. $S_x\propto E_z^3\, (1+z)^{-4}$ and $R_{200}\propto E_z^{-1}$). It is important to underscore that the cosmological parameters that we can actually constrain are those entering in the diameter distance and $E(z)$, that is $\Omega_m, \Omega_{\Lambda}, w(z)$.

In Fig. \ref{T-z} (right), we present the sensitivity of $\widetilde{EM}(x)$ at a fixed radius $\tilde x$ to the cosmological parameters $\Omega_m$ and $w$. We point out that the $\widetilde{EM}(x)$ dependence upon the underlying cosmology is stronger for increasing redshifts and for increasing $\beta$, i.e. in the outer regions, where $\beta\sim1$ \citep{morandi2015} with respect to the interiors ($\beta\lesssim0.7 $). This provides both a global (i.e. in the normalization, via the factor $E_z^{-3}$) and radial dependence (via a combination of $E_z$ and $D_{\rm a}$) on the cosmological parameters (equation \ref{dweded2}). For comparison, we also present the dependence of the measured gas fraction $f_{\rm gas}=M_{\rm gas} / M_{\rm tot}\propto D_{\rm a}(H_{0},\Omega_{\rm m}, w)^{3/2}$ on the cosmology: note that the gas fraction is a cumulative quality, i.e. it does not provide a radial dependence on the cosmological parameters.

It also important to point out that this cosmological test is insensible to the value of the Hubble constant, since we simply require that $\widetilde{EM}(x)$ is constant with redshift in the correct cosmology\footnote{More in general, we will explicitly allow for temperature and redshift dependence of $\widetilde{EM}(x)$, capturing possible departures from self-similarity in the normalization (via a temperature dependence of the baryon content) and evolution with redshift (via a redshift dependence of the baryon content) of $\widetilde{EM}(x)$. These departures from a strong self-similar model might arise due to the impact of non-gravitational processes (e.g. galaxy formation and feedback), leading to a generalized scenario known as weak self-similar model (Section \ref{coerg5uded2}).}, whose parameters must be determined a posteriori. In other words, the absolute normalization of $\widetilde{EM}(x)$ (which depends on the value of $H_0$) is a nuisance parameter in our analysis, being not of immediate cosmological interest. This is different, for example, from the standard hot gas mass fraction cosmological test, where relative hot gas fraction $f_{\rm gas}$ (whose value depends upon the Hubble constant) should be representative of the cosmic value $\Omega_b/\Omega_m$.

We point out that the dependence of $\widetilde{EM}(x)$ on the underlying cosmology is stronger than for $f_{\rm gas}$, and it is more pronounced as we move towards the outer volumes (for increasing $\beta$), providing both a global and radial dependence upon the cosmological parameters. Compared to other techniques (e.g. $f_{\rm gas}$, mass function), the proposed method has also the major advantage that it is more direct, i.e. based on observed quantities (the $EM$) on the plane of the sky, providing an independent way (and hinging on different assumptions) with respect to the former approaches. Unlike the gas fraction or the mass function methods, no assumption of hydrostatic equilibrium and three-dimensional geometry (since we do not infer the three-dimensional gas and total mass), homogeneous gas distribution\footnote{The possible impact of inhomogeneities of the gas distribution (`clumpiness') will be discussed in Section \ref{sgdgerdg3344}.}, stellar baryon fraction or model of mass function (calibrated e.g. on $N$-body simulations) is required, nor completeness of the cluster sample: we only require similarity of the (rescaled) $EM$ profiles. This cosmological test is also relatively insensible to X-ray calibration systematics. Note that the absolute normalization of $\widetilde{EM}(x)$ can vary, since we parametrize the dependence of $f_{\rm gas}$ on mass and redshift (Section \ref{coerg5uded2}).

However, it is also important to underscore that gas fraction measurements are characterized by significantly smaller intrinsic scatter for individual clusters ($\sim10-20$ percent) than $\widetilde{EM}(x)$ ($\sim30-60$ percent). Gas fractions and mass function methods can provide constraints also on other cosmological parameters besides $\Omega_m, \Omega_{\Lambda}, w(z)$, including the Hubble constant $H_0$ and the baryon fraction $\Omega_b$, and $\sigma_8$, respectively \citep{vikhlinin2009c,mantz2014}. Moreover, observations are very challenging in cluster outskirts, since the X-ray surface brightness drops below the background level at large radii. Hence, a large cluster sample is required to robustly measure $\widetilde{EM}(x)$ in the outer regions e.g. via stacking \citep[][]{morandi2015}, to improve the statistical significance of the $\widetilde{EM}(x)$ measurements and to reduce the observed scatter of the $EM$ profiles.

Finally, since we are interested in exploiting the large cosmological information `buried' in the outskirts of clusters (see Fig. \ref{T-z}), in practice we do not use $\widetilde{EM}(x)$ for individual clusters, rather we stack the raw $EM$ of several clusters grouped in bins of redshifts. This allow us to measure $\widetilde{EM}(x)$ for the stacked profiles out to $R_{100}$ and in several bins of redshift. The correct cosmology is thus that for which the stacked profiles $\widetilde{EM}(x)$ at different bins of redshift and for $x\gesssim0.2$ are (weakly) self-similar (see Section \ref{coerg5uded2} for a definition of weak self-similarity).

\subsection{Temperature and redshift dependence of $f_{\rm gas}$: weak self-similarity}\label{coerg5uded2}
As can be seen from equation (\ref{dweded}), clusters of given temperature evolve with redshift in a very simple way provided that they are self-similar objects, e.g. with normalization $EM_0\propto E_z^3$ and boundary radius $R_{200}\propto E_z^{-1}$. Naively, the comparison of cluster $EM$ at different redshifts can be thus used as `standard candle' to constrain the cosmological parameters, being a sensitive proxy of the underlying cosmology.

However, equation (\ref{dweded2}) highlights also the dependence of the scaled $EM$ on the gas fraction. While at the first order of approximation we can assume that $f_{\rm gas}$ (in the correct cosmology) is constant, i.e. independent of the cluster temperature and redshift, the question arises how much this assumption is tenable. Ultimately, the potential source of systematics in the use of cluster as cosmological probe could be ascribed to the presence of baryons since, in practice, the cluster observables are customarily locked to the baryonic component rather than the invisible (and nearly-perfectly self-similar) DM component. Cluster baryons are indeed subject to more complex physical processes (e.g. galaxy formation and feedback) besides the gravity, which are currently uncertain \citep{voit2005a} and difficult to model in hydrodynamic simulations \citep{kravtsov2012}, and which might evolve with redshift \citep{ettori2006}. These uncertainties might be enclosed in the $f_{\rm gas}$ term. A temperature and/or redshift dependence of $f_{\rm gas}$ could indicate the impact of the baryon component at the top of the evolution of the cosmic expansion history which affect the cluster observables via the term $H_z$.

In this respect, hydrodynamic numerical simulations outline a picture where the cumulative gas fraction is mildly increasing with system temperature when considered over a temperature (or mass) range extending from groups to massive clusters \citep[e.g.][]{young2011}. However, while the gas fraction in groups and low-mass systems implies an increasing trend of the cumulative gas mass fraction with temperature \citep{sun2009}, it is less clear whether such a trend holds for relatively massive objects. Measurements by \cite{vikhlinin2006} and \cite{allen2008} (both limited to the interiors of clusters) indicate that the gas fraction is independent of cluster temperature for objects with $kT > 5$ keV. In our previous work \citep{morandi2015} we did not find evidence of dependence of $f_{\rm gas}$ (in the outskirts) on the global temperature, which is probability due to the fact that most of the clusters (264 out of 320) in our sample have temperatures greater than 5 keV.

As for the redshift dependence, customarily measurements of the apparent evolution (or lack of) of the cluster X-ray gas mass fraction have been also used to probe the acceleration of the Universe, through the dependence of $f_{\rm gas}$ on the assumed distances to the clusters \citep{ettori2009}. In other words, the underlying idea is that the gas mass fraction is independent of the redshift in the `correct' cosmology. Observationally, our work \citep{morandi2015} provided for the first time a test of this assumption, indicating that $f_{\rm gas}$ is independent of the redshift. 

We caution the reader that the aforementioned findings on $f_{\rm gas}$ are somewhat cosmology dependent, while the cosmological parameters themselves are ultimately the quantity we are interested in. In order to consistently address any possible temperature and redshift dependence of $f_{\rm gas}$ and its ramification on the desired cosmological parameters, we examine this issue in a cosmology-independent way by incorporating a power-law with temperature and linear with redshift dependence into the model. We then marginalize over the slopes of this relation simultaneously with the cosmological parameters. Thus, uncertainties in the evolution of the expansion history due any possible dependence of $f_{\rm gas}$ on mass (or temperature) and redshift can be self-consistently accounted for. 

Hereafter, we shall use the following parametrization of the gas fraction:
\begin{equation}
f_{\rm gas}\propto (1+\alpha z){\left(\frac{T_X}{\rm 7\, keV}\right)}^{\delta}.
\label{eqn:fgas}
\end{equation}
Note that the normalization of $f_{\rm gas}$ is irrelevant for our purpose (equation \ref{dweded2}). We also point out that in equation (\ref{eqn:fgas}) we refer to a dependence of the gas fraction upon $T_X$ and $z$, rather than e.g. on the mass, the former two quantities being cosmology-independent\footnote{Strictly speaking, since the global temperature is calculated in the radial range $0.15-1\; R_{500}$ (e.g. Section \ref{swe222sim}), $R_{500}$ being cosmology dependent, $T_X$ might slightly depend on the assumed cosmology. However, in practice, this dependence is substantially negligible.}. We shall refer to $\alpha$ and $\delta$ as the {\em astrophysical} parameters, to distinguish them from the {\em cosmological} parameters.

By combining equations (\ref{dweded2}) and (\ref{eqn:fgas}), we can outline the dependence of $\widetilde {EM}(x)$ on both the astrophysical and cosmological parameters: 
\begin{equation}
\widetilde {EM}(x)\! \propto \!  (1+\alpha z)^{-2} {\left(\frac{T_X}{\rm 7\, keV}\right)}^{\!\!-1/2-2\delta}\!\!\! \!\!\!E_z^{-3+(1-6\beta)}\; D_{\rm a}^{1-6\beta}
\label{dweded2efw}
\end{equation}
with $x=R/R_{200}$ and $\beta=\beta(x)$. The expected evolution in the normalization of the scaling relations, which is solely due to the changing density of the Universe with redshift in the {\it strong} self-similar model, it is thus described in this {\it weak} self-similar model, capturing effects of non-gravitational processes at play in clusters \citep[][]{morandi2007b}. Therefore, different from \citet[][whose results hinge on the assumption of strong self-similarity]{arnaud2002}, we only require {\it weak self-similarity} of the (rescaled) $EM$ profiles, that is the shape of $\widetilde{EM}(x)$ is universal, explicitly allowing for temperature and redshift dependence of $f_{\rm gas}$.

Since the amount and nature of dark energy are poorly constrained by local clusters (equations \ref{eq:ez}), and in light of the dependence of the observables quantity $\widetilde {EM}(x)$ on the underlying cosmological parameters (equation \ref{dweded2}), we underscore that we need a large sample of clusters: (i) spanning a wide range of redshift; (ii) where $\widetilde {EM}(x)$ can be inferred out to large radii via e.g. stacking (since we are interested in exploiting the large cosmological information `buried' in the outskirts of clusters, see Fig. \ref{T-z}); (iii) where we have exquisite knowledge of the systematics due to the X-ray analysis, given the difficulties to infer X-ray signal in the outskirts; and (iv) with the aid of complementary data sets (e.g. CMB) and priors on the astrophysical parameters (e.g. from simulations), in order to break the degeneracy among the parameters. 

In particular, concerning the last point, we emphasize the importance of external data sets in order to have a bound on both the parameters of the underlying cosmology and the evolution of $f_{\rm gas}$ from theory or simulations. Indeed, e.g. an apparent evolution (or lack of) in $\widetilde {EM}$ as measured from X-ray data could be ascribed to the cosmic expansion history and EoS of dark energy (equation \ref{dweded2}); but at the same time it could be related to the evolution of the overall hot fraction budget $f_{\rm gas}$ due to e.g. non-gravitational processes (equation \ref{eqn:fgas}). Hence, the real challenge is to measure and distinguish these two evolutions which can be attributed to both cosmological and astrophysical processes.

In the following section we will present our X-ray sample, including particulars on the temperature and redshift of the clusters, and details on the data analysis.

\section{Data sets and analysis}\label{swe222sim}

\subsection{X-ray sample and analysis}\label{swe22brim}
This X-ray data analysis project represents a follow-up of our previous works on studying cluster physical properties, including $EM$, gas density and gas fraction in the outskirts \citep{morandi2015}. We have shown that {\em Chandra} can accurately measure the physical properties of clusters in the virialization region by stacking the emission measure profiles $EM\propto \int n_e^2\, dl$ of the cluster sample to detect a signal out to and beyond $R_{200}$.

\begin{figure}
\begin{center}
\hbox{
\includegraphics[scale=0.42]{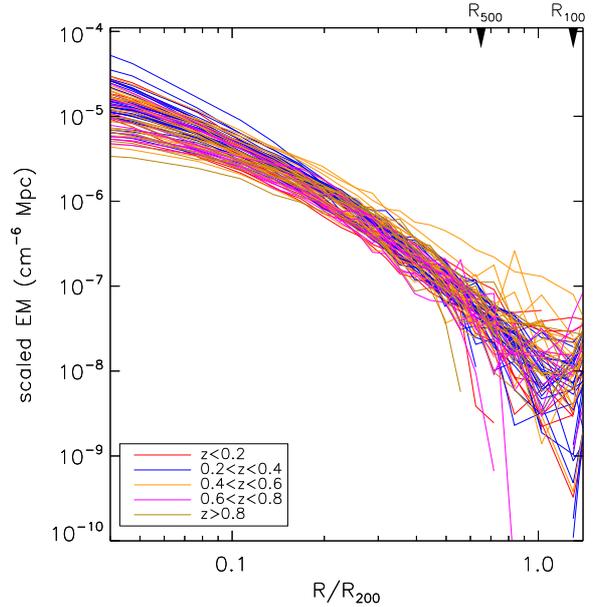}
}
\caption{The {\em Chandra} emission measure profiles of all clusters in our sample, re-scaled according to the self-similar model (Section \ref{selfsim}). A straight line is drawn between the data points for each cluster for a better visualization of the profiles. The curves are colour-coded by redshift bins (Section \ref{rjmcmc1}), and we assumed a reference flat $\Lambda$CDM model, with matter density parameter $\Omega_{\rm m}=0.3$ and $H_{0}=70 \,{\rm km\; s^{-1}\; Mpc^{-1}}$ in the self-similar scaling. Note the remarkable similarity of the profiles outside the cluster cores, with high-$z$ cluster showing similarity in shape and normalization with nearby sources. For each redshift interval, we sort the clusters in redshift and stacked the profiles in bins of four sources to reduce the scatter and improve the readability of the figure.}
\label{T-zegry5}
\end{center}
\end{figure}

The sample is composed of 320 clusters from the {\em Chandra} archive \citep{morandi2015}, with redshift range $z=0.056-1.24$ (the median redshift is $\sim0.4$) and temperatures $kT\gesssim 3$ keV (the median temperature is $\sim 7$ keV). Cluster $EM$ profiles are presented in Fig. \ref{T-zegry5}. These sources have been observed with the I-array of the Advanced CCD Imaging Spectrometer (ACIS-I) and they have been selected because the X-ray observations encompass $R_{200}$ of each cluster. Thus, these sources are ideal for observations of cluster outskirts. Moreover, source-free regions of the cluster observations are always present, allowing measurements of the local background which is fundamental for robust measurements of the physical parameters in the outskirts.

Although the sample was selected based on the quality of the existing observations and hence might be subject to selection effects, for the purpose of this work we did not require that the sample be complete, given the high level of self-similarity of the cluster outskirts. A posteriori, we verified that our working hypothesis (self-similarity) is robust towards systematics by considering cluster sub-samples based on redshift, temperature, relaxation state, observing time, and repeating our analysis (Section \ref{sgnegneg53}). We emphasize again that our analysis concerns primarily relatively hot clusters, for which non-gravitational effects like pre-heating should have a marginal impact, and the measured slopes of scaling relations should be close to the self-similar scaling laws. Departures from these theoretical expectations are discussed in Section \ref{coerg5uded2}.

Here we summarize the most relevant aspects of the X-ray analysis. For further details we refer to \citet{morandi2015}. 

We have performed a spectral analysis in order to infer the global spectroscopic temperature $T_X$ via the \texttt{XSPEC} package \citep[][version 12.8.2]{1996ASPC..101...17A}. We adopted the APEC emissivity model \citep{foster2012}, the AtomDB (version 2.0.2) data base of atomic data and the solar abundance ratios from \cite{asplund2009}. We also used the Tuebingen-Boulder absorption model (\texttt{tbabs}) for X-ray absorption by the interstellar medium, and we fixed the hydrogen column density $N_H$ to the Galactic value by using the Leiden/Argentine/Bonn (LAB) HI-survey \citep{kalberla2005}. Spectra have been thus extracted from a circular region in the radial range $0.15-0.75\; R_{500}$ around the cluster centroid, once the contaminating point sources are masked. The only free parameters in the spectral fit are the temperature, the normalization and the metallicity, while we fix the redshift to the value obtained from optical spectroscopy. The background spectra have been extracted from regions of the same exposure for the ACIS-I observations, for which we always have some areas ($\gesssim R_{100}$) free from source emission. We then corrected $T_X$ to infer a value in the radial range $0.15-1\; R_{500}$ via mock simulations. This correction factor is $0.96\pm0.02$. 

We then measure the $EM$ profile $EM\propto \int n_e^2\, dl$ from the X-ray images. The radial $EM$ profile is derived with the exposure correction. We applied a direct subtraction of the cosmic X-ray (CXB)+particle+readout artefact backgrounds. For the particle background modelling, we use the scaled stowed background. In order to measure the CXB, we considered the region where the CXB is more dominant than the cluster emission, which can be determined from the flattened portion at the outer region of the surface brightness profile. We also accounted for the readout background using the {\tt make\_readout\_bg} routine. Finally, in order to account for the vignetting, we modelled the soft CXB component by an absorbed power law with index 1.4 and two thermal components at zero redshift, one unabsorbed component with a fixed temperature of 0.1 keV and another absorbed component with a temperature derived from spectral fits \citep[$\sim0.25$ keV,][]{sun2009}.

$EM(x)$ has been then rescaled according to the self-similar model (equation \ref{dweded2}), in order to infer a renormalized value $\widetilde{EM}(x)$. For each cosmological set, we calculated the median distribution of the stacked profile $\widetilde{EM}(x)$, which has been then deprojected in order to infer the gas density profiles, the gas mass, and the X-ray Compton parameter $Y_X=M_{\rm gas}\, T_{X}$ \citep{kravtsov2006}. Thus, we infer $R_{500}$ from the observables quantities via the $Y_{X}-M$ relation measured for the \citet{sun2009} sample of clusters. We then calculated $R_{\Delta}$ ($\Delta=200,100$) from the known value of $R_{500}$ by assuming an NFW distribution with parameters $(c,R_{200})$, with the concentration parameter $c$ from the $c-M$ relation of \cite{dutton2014}. This provides an estimate of $R_{\Delta}$ for given value of $R_{500}$. For our sample we have $R_{500}=(0.65\pm0.01) R_{200}$ and $R_{100}=(1.36\pm0.01) R_{200}$.

Note that in our work we have considered all the physical quantities at fixed overdensity (e.g. $\Delta=200$) with respect to the critical density $\rho_{\rm c,z}$. We will discuss how this definition is robust towards other definition of virial radius.

\subsection{External Cosmological Data Sets}\label{swe22brim2}
Our analysis will make use of complementary data sets as well in constraining both astrophysical and cosmological parameters.
In particular, we exploit the `Union2.1' SNIa compilation \citep{suzuki2012} in combination with {\em Planck} CMB data \citep{planck2015a} and {\em Wilkinson Microwave Anisotropy} (WMAP) polarization measurements \citep{bennett2013}. For these data sets, likelihood functions are provided by the Planck collaboration \citep{planck2015a}. We inferred marginal probability distributions of the subset of cosmological parameters (e.g. $\Omega_m$, $\Omega_{\Lambda}$ and $w(z)$) from these likelihood functions for the given cosmological model.

Note that the likelihood for the {\em Planck} data is not computed using the covariance matrix for the cosmological parameters of interests, rather we explore the entire multidimensional grid of the cosmological parameters from the Markov chain simulations. In this way, we recover the entire information from the location and relative amplitudes of the peaks in the CMB power spectrum and SNIa constraints.

\subsection{Priors on $f_{\rm gas}$ evolution from hydrodynamic simulations}\label{sw555bsim}

Hydrodynamic simulations of cluster formation predict a scenario where the gas mass fraction at intermediate to large cluster radii should have a small cluster-to-cluster scatter, with no evolution with redshift \citep{nagai2007a,young2011,battaglia2013,planelles2014}. Thanks to the improvements in the models of baryonic physics, in particular modelling of the effects of feedback from active galactic nuclei (AGNs), hydrodynamic simulations have become able to better recover the gas fraction in galaxy clusters and its evolution with cosmic time.

Specifically, we consider the recent smoothed particle hydrodynamics (SPH) simulations of \cite{planelles2014}, which carried out one set of non-radiative (NR) simulations, and two sets of simulations including radiative cooling, star formation and feedback from SNe (CSF), one of which also accounting for the effect of feedback from AGNs. They evaluated the gas fraction and depletion from these simulations at different overdensities, studying their evolution (or lack of) with redshifts according to equation (\ref{eqn:fgas}).

Combining these predictions with measurements of the X-ray $EM$, as well as from constraints on the cosmology from external data sets, provides a robust method to constrain both cosmological and astrophysical parameters at play in our analysis, greatly reducing their degeneracy. We will primarily focus on the AGN-simulation results of \cite[][see their Table 3]{planelles2014}, incorporating their results as Gaussian priors on $\alpha$ (the parameter describing the evolution with $z$ of $f_{\rm gas}$, e.g. equation \ref{eqn:fgas}). However, we will also investigate the impact of their different set of simulation (NR and CSF) on the evolution of $f_{\rm gas}$ and hence the desired cosmological parameters. Finally, we will also gauge the impact of a different set of simulations \citep{battaglia2013}.

\section{Model fitting and selection}\label{rjmcmc1}
A Bayesian statistics can be used to constrain both cosmological parameters ($\Omega_m$, $\Omega_{\Lambda}$, $\Omega_k$ and $w(z)$) inference and model selection. Concerning the former point, we are primarily interested in constraining the EoS of dark energy. Concerning the model selection, we shall determine which cosmological model $\Lambda$CDM, $w$CDM and $w_z$CDM and definition of `virial' radius are more likely given the data. 

Concerning the model fitting, we used the Metropolis-Hastings Markov chain Monte Carlo (MCMC) algorithm for sampling from multi-dimensional distribution. In this way, we cycle through all the parameters $\btheta$ of the cosmological model ${\mathcal{H}_{i}}$ (e.g. $\Lambda$CDM, $w_z$CDM etc.), updating each from conditional distributions, in order to achieve convergence to a stationary posterior distribution $P(\btheta|\vect{D},\mathcal{H}_i)$ for a given data set $\vect{D}$. This level of inference is repeated for different cosmological models $\mathcal{H}_i$.

For each cosmological model, our merit function is represented by $\widetilde{EM}$. In practice, since we are interested in exploiting the large cosmological information `buried' in the outskirts of clusters (see Fig. \ref{T-z}), we stack the $EM$ of several clusters grouped into bins of redshifts in order to measure $\widetilde{EM}(x_j;z_i)$ for the stacked profiles out to $R_{100}$, in the $j$th radial annulus ($x_j=R_j/R_{200}$) and in the $i$th bin of redshift. We divided our 320 X-ray clusters into five redshift bins ($z_1$:$z$$<$0.2; $z_2$:0.2$<$$z$$<$0.4; $z_3$:0.4$<$$z$$<$0.6; $z_4$:0.6$<$$z$$<$0.8; $z_5$:$z$$>$0.8), in order to have $\gesssim50$ clusters per bin. This rebinning ensures a statistically significant detection of the X-ray signal in the outskirts at each redshift $z_i$, $i=1,...,5$. The correct cosmology is thus that for which the stacked profiles $\widetilde{EM}(x_j; z_i)$ at redshifts $z_i$ and for $x\gesssim0.2$ are (weakly) self-similar (see Section \ref{coerg5uded2} for a definition of weak self-similarity).

Hence the likelihood (the proposal distribution) ${\mathcal{L}}\propto \exp(-\chi^2/2)$, and $\chi^2$ reads:
\begin{equation}\label{chi2wwe}
\chi^2= \sum_{i>1,j} {\frac{{ (\widetilde{EM}(x_j;z_i)-\widetilde{EM}(x_j;z_1))}^2 }{ \sigma_{j,i}^2 + \sigma_{j,1}^2 }}\
\end{equation}
where $\sigma_{j,i}$ is the uncertainty (total scatter) of $\widetilde{EM}(x_j;z_i)$. This posterior probability allows us to perform a quantitative comparison of the scaled $EM$ profiles of distant clusters with a local reference profile derived from hot nearby clusters (at $z\lesssim0.2$). Note that the amount and nature of dark energy and matter density have a negligible impact on the scaled $EM$ at the lowest redshifts.

The dependence of the stacked emission measure $\widetilde{EM}(x_j;z_i)$ on the cosmological and astrophysical parameters is outlined in equation (\ref{dweded2efw}). We obtain parameter constraints using the likelihood function computed for each proposed state $\btheta_k$ (e.g. $\btheta=\{\Omega_m, \Omega_{\Lambda},w_0,w_a, \alpha,\delta\}$) of cosmological+astrophysical parameters affecting cluster observables (and with/without priors from external data sets and priors on $\alpha$ from hydrodynamic simulations) in the MCMC analysis. This is computationally demanding since we need to re-calculate the stacked emission measure $\widetilde{EM}(x_j;z_i)$ for each proposed state $\btheta_k$ in all the radial and redshift bin, and we describe our approach below. We calculated the raw X-ray surface brightness $S_x(\theta)$ profile (expressed in counts s$^{-1}$ arcmin$^{-2}$, with $\theta=x\,R_{200}/D_a$) of each cluster, which can be regarded as {\it cosmological independent}, following a predefined binning in units of $R_{200}$ (calculated in a reference cosmology) common to all clusters. We then performed the median to compute stacked $\widetilde{EM}(x_j;z_i)$ profiles in each bin of redshift. For each proposed state $\btheta_k$, we interpolated the individual surface brightness profiles $S_x(\theta)$ following a new binning in units of $R_{200}$ (calculated in the proposed cosmology), and compute new stacked $\widetilde{EM}(x_j;z_i)$ profiles which reflect the proposed parameters $\btheta_k$ (equation \ref{dweded2efw}). For each proposed state $\btheta_k$ the errors $\sigma_{j,i}$ (total scatter) on the interpolated points were propagated to the stacked profiles according to the Monte Carlo (MC) approach described in \cite{morandi2015}.

Moreover, in order to seek for an improved proposal distribution ${\mathcal{L}}(\btheta,\mathbfit{C})$, we perform an adaptive MCMC, which asks the code to automatically `learn' better parameter values `on the fly', i.e. by capturing the covariance matrix $\mathbfit{C}$ of the parameters while the algorithm runs. We then generated proposed parameters with multinormal probability distribution ${\mathcal{L}}(\btheta,\mathbfit{C})$, i.e. with expectation value $\btheta$ and covariance $\mathbfit{C}$, by performing a Cholesky decomposition of $\mathbfit{C}$. We refer to the classical statistical books for a detailed discussion about this methodology \citep{mackay2003}.

In addition to a complete statistical analysis of the chains, we performed convergence tests, e.g. the Gelman \& Rubin R statistics \citep{gelman1992}. The typical number of MCMC iterations is $\sim10^{6}$ for the post-convergence, and a few $\sim10^{4}$ for the burn-in period.

Finally, we empirically verified that equation (\ref{chi2wwe}) gives an unbiased $\chi^2$ via the following test. From the rescaled $EM$ averaged on the whole sample, we created mock $\widetilde{EM}(x_j;z_i)$ profiles for each redshift bin $z_i$, with uncertainty $\sigma_{j,i}$ and for an assumed cosmology. We then repeated the cosmological analysis, by inferring new $\widetilde{EM}(x_j;z_i)$ profiles which reflect the proposed parameters $\btheta_k$. We thus verified agreement between input and output cosmological parameters. %within the "MCMC accuracy" inversely proportional to the square root of the MCMC sample size

Concerning the model comparison, we point out that a Bayesian model selection will be possible since our priors on the cosmological and astrophysical parameters are informative, defining statistical information about fitting variables. Our analysis indeed hinges on the constraints on the cosmological parameters due to our assumption of (weak) self-similarity of the stacked emission measure $\widetilde{EM}(x)$ (the `likelihood' $P(\vect{D}|\btheta,\mathcal{H}_i)$), but also on priors $P(\btheta | \mathcal{H}_i)$ on the cosmological parameters from complementary data sets $\vect{D}$ (e.g. CMB, SNe) and priors on the astrophysical parameter $\alpha$ (Section \ref{coerg5uded2}) from hydrodynamic simulations. 

In particular, we use the evidence-based criterion. Assuming that we choose to assign equal priors $P(\mathcal{H}_i)$ to the alternative cosmological models $\mathcal{H}_i$ (e.g. $\mathcal{H}=w_z CDM$), the models are ranked by evaluating the evidence ratios: 
\begin{equation}\label{bayes4r}
K=P(\mathcal{H}_i|\vect{D}) /P(\mathcal{H}_j|\vect{D})= P(\vect{D}|\mathcal{H}_i)/P(\vect{D}|\mathcal{H}_j) 
\end{equation}
where $K$ is called Bayesian factor. A value of $K>1$ means that ${H}_i$ is more strongly supported by data under consideration than ${H}_j$. For the interpretation of the Bayesian factor $K$ we refer to the scale of \cite{kass1995}.

\section{Constraints on the EoS of dark energy: results}\label{swefffsim}

This section presents the cosmological constraints measured through our analysis of the cluster stacked $EM$. We will start by presenting the constraints arising in a flat (and non-flat) $\Lambda$CDM and $w$CDM models, in a classical Bayesian framework, where the cosmological model $\mathcal{H}_i$ is known a priori. This will allow us to have a feeling on the quality and degeneracies of these constraints, compared to other constraints from observables pertaining to the expansion history and structure growth. 

We will then tackle more complex cosmological models using the cluster data, as well as via external cosmological probes and constraints on the evolution of $f_{\rm gas}$ from hydrodynamic simulations used as priors in our Bayesian analysis. This will allow us to constrain simultaneously cosmological models and astrophysical+cosmological parameters. When combining data sets, we will consider combinations which cumulatively include cluster constraints, {\em Planck} CMB or complementary data sets, as well as priors on the $f_{\rm gas}$ evolution, in order to gauge the impact of different assumptions and systematics, and remove the degeneracy among the fitting parameters.

The impact of several sources of systematics in our analysis, including biases due to the modelling of the background, the presence of outlier measurements, selection effects, inhomogeneities of the gas distribution and cosmic filaments, will be discussed in Section \ref{sgnegneg53}.

\subsection{Constraints on non-flat $\Lambda$CDM models}\label{swefrre}

\begin{figure}
\begin{center}
\hbox{
\includegraphics[scale=0.39]{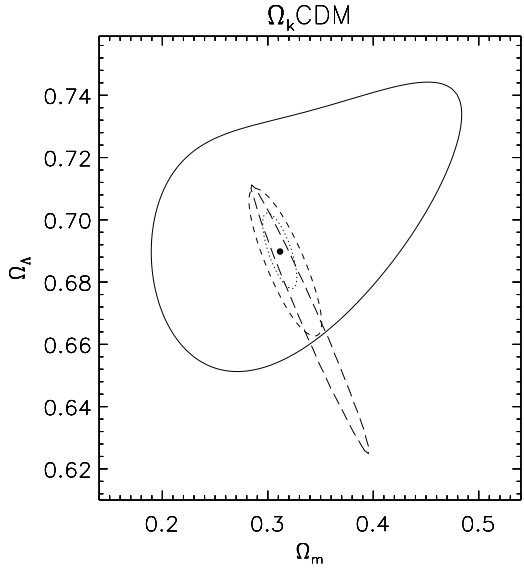}
}
\caption{2D marginalized posterior distributions for $\Omega_m$ and $\Omega_{\Lambda}$ in a non-flat $\Lambda$CDM model ($\Omega_k$CDM). We show constraints arising from our X-ray-only analysis of the stacked emission measure $\widetilde{EM}(x)$ in the whole {\em Chandra} sample (solid line); and by cumulatively adding the likelihood from the Planck+WP+Union2.1 data sets (dashed line) and constraints on the evolution of $f_{\rm gas}$ from hydrodynamic simulations (dotted line) as priors. The thick long-dashed line shows the marginalized posterior distribution only from the Planck+WP+Union2.1 data sets. The contours are at the 1-$\sigma$ confidence level, while the expectation value is represented by a point and refers to the combination of all the data sets. \protect\\Notes on the complementary data sets: `{\em Planck}' refers to the CMB analysis of \protect\cite{planck2015a}; `Union2.1' refers to the SNIa compilation of \protect\cite{suzuki2012}; `WP' refers to the WMAP polarization measurements \protect\citep{bennett2013}; the constraints on the evolution of $f_{\rm gas}$ from hydrodynamic simulations refer to the results of \protect\citet{planelles2014}.}
\label{T-zedf3452}
\end{center}
\end{figure}

In this subsection, we investigate constraints on the parameter $\Omega_k$, where for non-flat $\Lambda$CDM models (hereafter $\Omega_k$CDM) $\Omega_k = 1-\Omega_{m} - \Omega_{\Lambda}$. The simple assumption of large-scale homogeneity and isotropy leads to the Friedmann-Robertson-Walker (FRW) metric. For FRW models $\Omega_k>0$ would correspond to negatively-curved 3D-geometries (or open-Universes), while $\Omega_k<0$ to positively curved 3D geometries (or closed Universes). The baseline $\Lambda$CDM cosmology hinges on an FRW metric with a flat curvature ($\Omega_k=0$), which has been proved to be an accurate description of our Universe as constrained by CMB measurements \citep{planck2015a}. In this respect, our analysis provides an independent method to test empirically this assumption.

Our constraints obtained from the full X-ray data set (with/without standard priors and external data sets) are shown as contours at the 1-$\sigma$ confidence level in Fig. \ref{T-zedf3452}. For this model, we have two cosmological and two astrophysical parameters, i.e. $\btheta=\{\Omega_m, \Omega_{\Lambda}, \alpha,\delta\}$. We obtain $\Omega_m = 0.350 \pm 0.116$, $\Omega_{\Lambda}= 0.708\pm 0.037$ and $\Omega_k=-0.058\pm 0.144$ for the case without external data sets and/or priors, with a small (positive) correlation between the two parameters, as can be seen in the figure. Also shown are independent constraints from Planck+WP+Union2.1 \citep{suzuki2012,bennett2013,planck2015a}. While the two independent data sets are in good agreement, it is interesting to point out that the degeneracy between $\Omega_m$ and $\Omega_{\Lambda}$ in these data sets is nearly orthogonal, with the external data sets providing a more stringent constraint on the curvature $\Omega_k$ of the Universe. 

Combining all the data sets (without leveraging on additional priors), we obtain stringent constraints strongly pointing to a flat Universe: $\Omega_m = 0.318 \pm 0.042$ and $\Omega_{\Lambda}=0.690 \pm 0.019$, with $\Omega_k =-0.009 \pm0.046 $. If we add priors on the gas fraction evolution from hydrodynamic simulations we have $\Omega_m = 0.311 \pm 0.015$ and $\Omega_{\Lambda}=0.690 \pm 0.011$, with $\Omega_k =-0.001 \pm0.012 $. The spatial curvature of the Universe is thus found to be close to zero with an uncertainty of the order of 0.01, in agreement with the uncertainty found by \cite[][$\sim0.005$]{planck2015a}.

\subsection{Constraints on flat $\Lambda$CDM models}\label{swefffsim3e}

\begin{figure*}
\begin{center}
\hbox{
\includegraphics[scale=0.43]{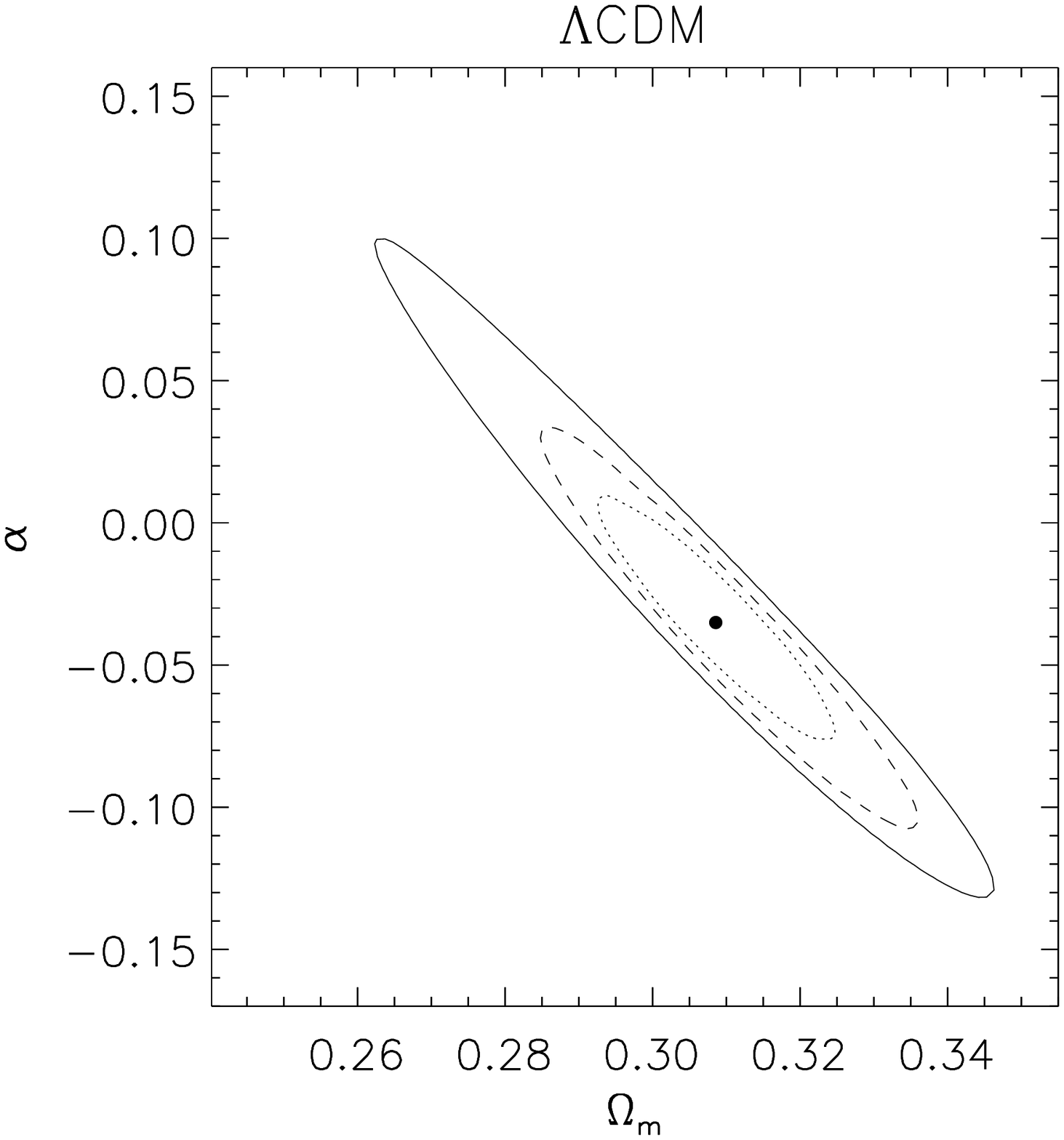}
\includegraphics[scale=0.43]{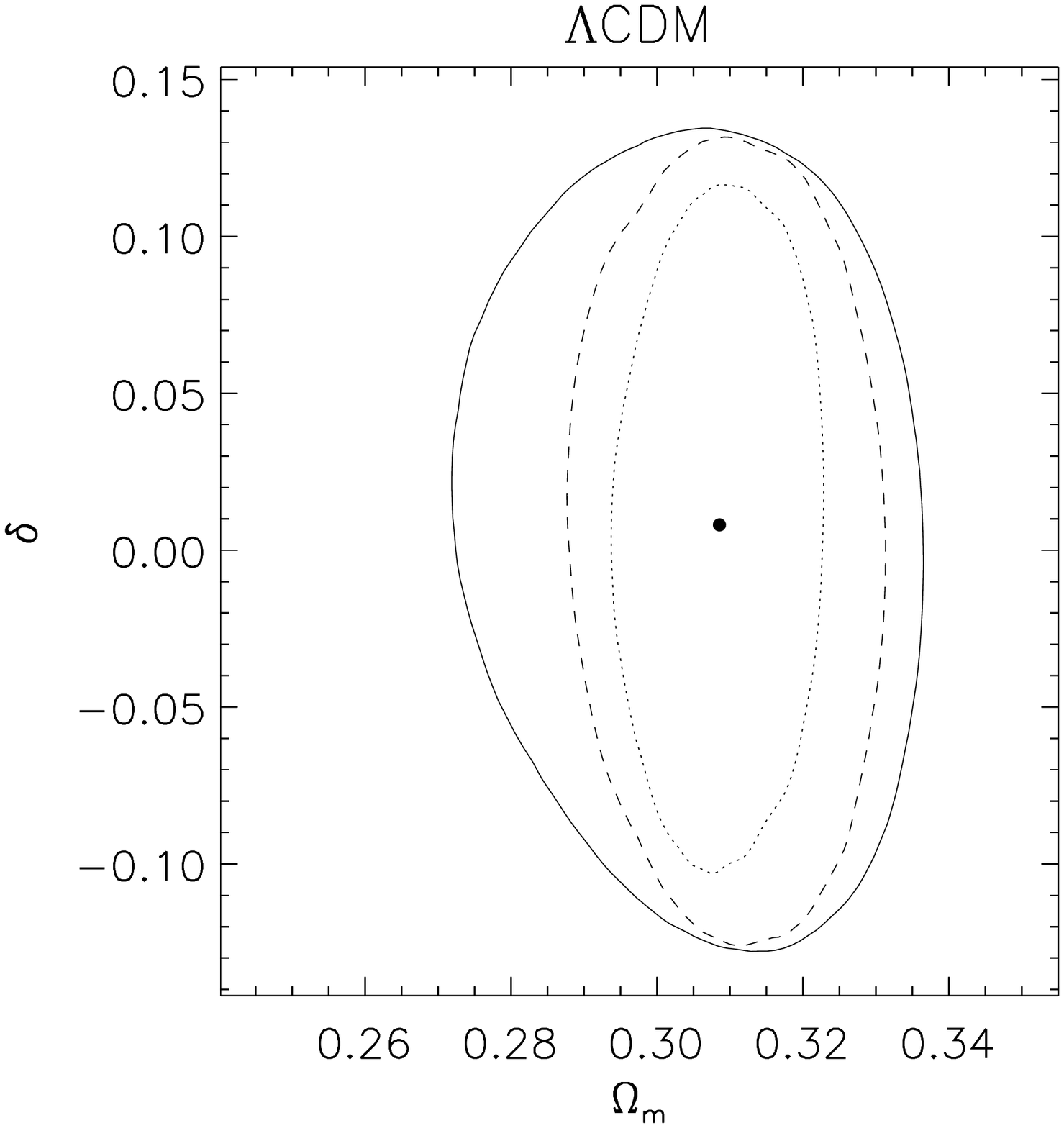}
}
\caption{Left-hand panel: 2D marginalized posterior distributions for $\Omega_m$ and $\alpha$ (the parameter describing the evolution with $z$ of $f_{\rm gas}$, e.g. equation \ref{eqn:fgas}) in a flat $\Lambda$CDM model. Right-hand panel: 2D marginalized posterior distributions for $\Omega_m$ and $\delta$ (the parameter describing the temperature dependence of $f_{\rm gas}$, e.g. equation \ref{eqn:fgas}) in a flat $\Lambda$CDM model. For all the panels, we show constraints arising from our X-ray-only analysis of the stacked emission measure $\widetilde{EM}(x)$ in the whole {\em Chandra} sample (solid line); and by cumulatively adding the likelihood from Planck+WP+Union2.1 data sets (dashed line) and constraints on the evolution of $f_{\rm gas}$ from hydrodynamic simulations (dotted line) as priors. The contours are at the 1-$\sigma$ confidence level, while the expectation values are represented by a point and refers to the combination of all the data sets. \protect\\Notes on the complementary data sets: `{\em Planck}' refers to the CMB analysis of \protect\citet{planck2015a}; `Union2.1' refers to the SNIa compilation of \protect\cite{suzuki2012}; `WP' refers to the WMAP polarization measurements \protect\citep{bennett2013}; the constraints on the evolution of $f_{\rm gas}$ from hydrodynamic simulations refer to the results of \protect\citet{planelles2014}.}
\label{T-z3452b}
\end{center}
\end{figure*}

For a flat $\Lambda$CDM models ($\Lambda$CDM), the constraints obtained from the full X-ray data set are shown in Fig. \ref{T-z3452b} as contours at the 1-$\sigma$ confidence level. For this model, we have one cosmological and two astrophysical parameters, i.e. $\btheta=\{\Omega_m, \alpha,\delta\}$. We emphasize a strong anti-correlation between $\Omega_m$ and $\alpha$. As we previously pointed out (Section \ref{coerg5uded2}), the apparent evolution with $z$ (or lack of) of the stacked emission measure $\widetilde {EM}$ as measured from X-ray data could be ascribed to both the cosmic expansion history of the Universe (via the parameter $\Omega_m$) and to the evolution parameter $\alpha$ of $f_{\rm gas}$ due to e.g. non-gravitational processes (equation \ref{dweded2efw}). With respect to the correct cosmology, an (apparent) positive evolution with $z$ of $\widetilde {EM}$ could be qualitatively achieved both by a smaller value of $\Omega_m$ (Fig. \ref{T-z}, right-hand panel) or with negative evolution of the hot gas fraction ($\alpha<0$). This translates into an intrinsic degeneracy between astrophysical and cosmological parameter, which limits our ability to infer the desired parameters from our X-ray-only analysis of $\widetilde{EM}(x)$ without external data sets and/or priors. We have $\Omega_m = 0.305\pm 0.022 $ and $\alpha=-0.021\pm 0.062$, consistent with no evolution of $f_{\rm gas}$. 

In Fig. \ref{T-z3452b} we can easily identify the role of using external data sets and/or priors on the hot gas fraction evolution in constraining the cosmological parameters. Clearly, it breaks the degeneracy mostly along the $\Omega_m$ axis. The best-fitting values and statistical uncertainties for $\Omega_m$ are very close to those derived from the external data sets. The cluster constraints on $\Omega_m$ combined with Planck+WP+Union2.1 data sets provide $\Omega_m = 0.310 \pm 0.014$ and $\alpha=-0.037 \pm0.039$, with a significant improvement in the statistical errors. For comparison, from CMB-only Planck temperature combined with Planck lensing, we have a matter density parameter $0.308 \pm 0.012$, outlining how the CMB constraints are more stringent for $\Lambda$CDM in constraining the matter density parameter with respect to those from X-ray-only data (without any additional prior/external data set). 

Adding the prior on $\alpha$ from hydrodynamic simulations leads to the CMB-free constraints on the matter density parameter $0.307\pm 0.011$, which is competitive in term of statistical uncertainties and in agreement with the CMB-only constraints. 

While the advantage of introducing theoretical priors for the gas fraction evolution significantly improves the parameters constraints, we caution the reader that the reliability of the theoretical predictions might deserve some prudence. Improved hydrodynamic numerical simulations with cooling, star formation and feedback become able to better recover the baryonic content of clusters \citep{battaglia2013}. However, systematic differences between different sets of hydrodynamic simulations (e.g. SPH or adaptive-mesh refinement codes, see \citealt{rasia2014}) and the impact of the particular baryonic physics implemented \citep{planelles2014} suggest that the use of these theoretical priors on $\alpha$ merits some caution. In order to gauge the impact of these systematic uncertainties in our recovered parameters, we used the results \cite{planelles2014}. These are based on one set of NR simulations, and two sets of simulations including radiative cooling, star formation and feedback from SNe (CSF), one of which also accounting for the effect of feedback from AGNs. While we usually refer to their AGN results as baseline prior, the use of the other predictions on $\alpha$ for CSF and NR simulations, which encompass the aforementioned systematic uncertainties in the physics prescriptions, can give us a sense of the impact of these biases in our analysis (when these priors are included). By incorporating these different predictions on $\alpha$ as priors in our Bayesian MCMC analysis performed on mock data sets (see Section \ref{sggg53} for further details on the mocks), we conservatively estimate that the bias (downwards) on the matter density parameter is $\lesssim0.5\%$ and $\lesssim1\%$ for the CSF and NR hydrodynamic simulations, respectively. This must be compared to statistical uncertainties $\sim5\%$ (with external data set). The level of bias from NR simulations probably overestimates the true systematics, since NR hydrodynamic simulations likely do not capture the correct physics in the ICM. We shall therefore focus on the two cases AGN and CSF in order to estimate these systematics. Similar conclusions hold by considering the predictions on the gas fraction evolution of \cite{battaglia2013}.

Thus, by combining all the data sets and priors, we have $\Omega_m = 0.309 \pm 0.009$ and $\alpha=-0.034 \pm0.024$, with the latter quantity still consistent with zero (i.e. no evolution of $f_{\rm gas}$). We also quote the value of $\delta$ (the parameter describing the temperature dependence of $f_{\rm gas}$, e.g. equation \ref{eqn:fgas}): $\delta=0.006\pm 0.072$, with no evidence for a trend with cluster temperature (or mass) of $f_{\rm gas}$. Note that, unlike the $f_{\rm gas}-z$ relation, marginalizing over an $f_{\rm gas}-kT$ slope has a negligible effect on our cosmological constraints (e.g. in their errors), as we verified by fixing the temperature dependence to zero. In \cite{morandi2015} we also found a lack of dependence of the gas fraction on the cluster temperatures (in a cosmology dependent way). This is likely explained by the relatively narrow temperature range in our sample ($\gesssim$ 3 keV), such that the systems in our sample should show  negligible dependence of the physical properties (e.g. the emission measure) on the global temperature. Indeed most of the clusters (264 out of 320) in our sample have temperatures greater than 5 keV. While the gas fraction in groups and low-mass systems implies an increasing trend of the cumulative gas mass fraction with temperature \citep{sun2009}, it is less clear whether such a trend holds for relatively massive objects. Measurements by \cite{vikhlinin2006} and \cite{allen2008} (limited to the interiors of clusters) indicate that the gas fraction is independent of cluster temperature, for objects with $kT > 5$ keV, in agreement with our findings. This trend is also confirmed by the level of the intrinsic scatter of $f_{\rm gas}$ ($\sim15\%$ at $R_{500}$, $\sim25\%$ at $R_{200}$), which does not appear to depend on the selected subsamples.

Finally, we compare our results with \cite{arnaud2002}. They study the surface brightness profiles of a sample of 25 distant and hot clusters, observed with ROSAT, with published temperatures from ASCA. They performed a quantitative comparison of the scaled $EM$ profiles of these distant clusters with a local reference profiles derived from the sample of 15 hot nearby clusters. Using (strong) self-similar evolution model, they put stringent constraints on the matter density of a $\Lambda$CDM model, obtaining $\Omega_m=0.40^{+0.15}_{-0.12}$ at 90\% confidence level.

\subsection{Constraints on flat $w$CDM models}\label{swefffsim3ed}

\begin{figure*}
\begin{center}
\hbox{
\includegraphics[scale=0.43]{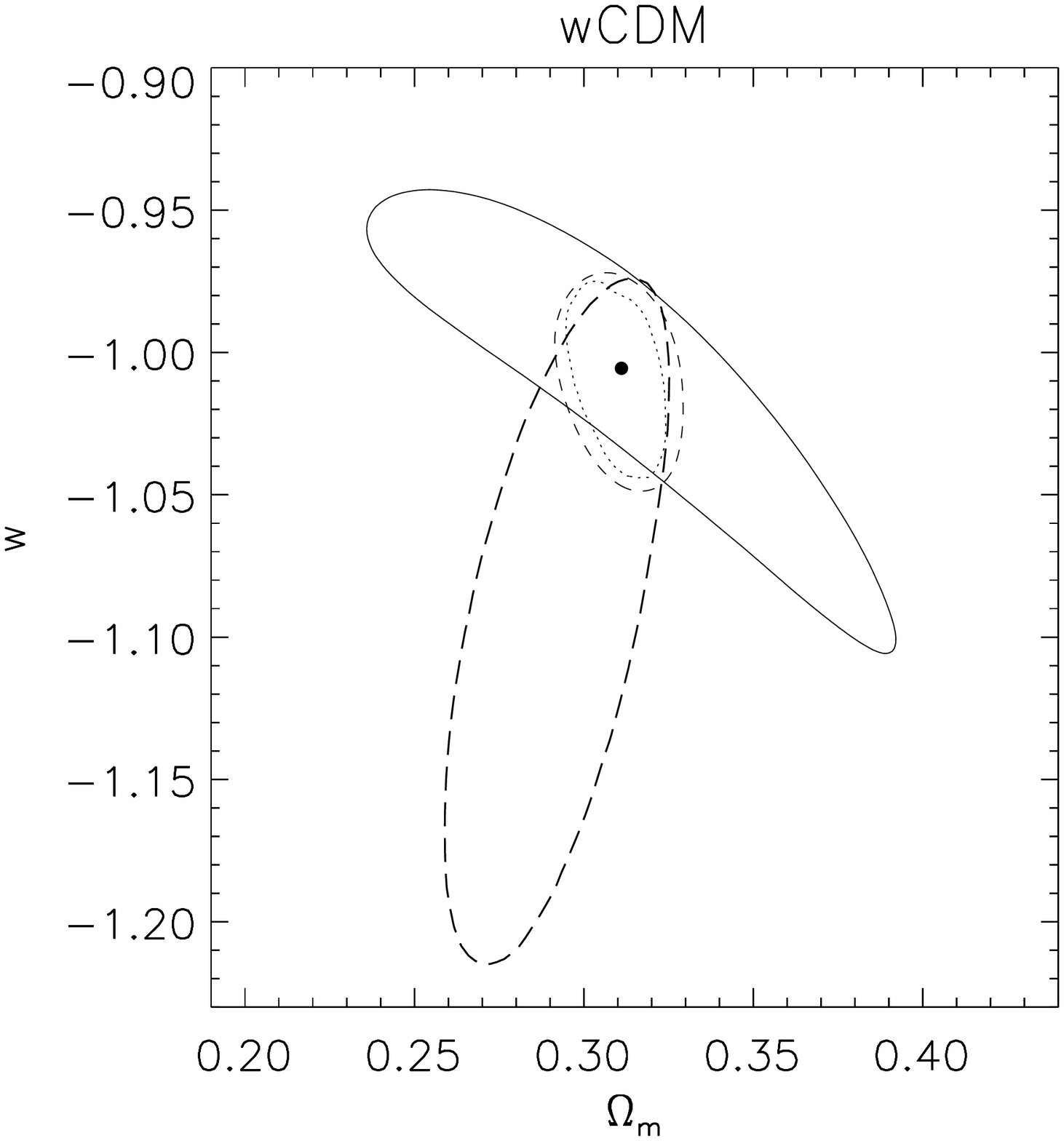}
\includegraphics[scale=0.43]{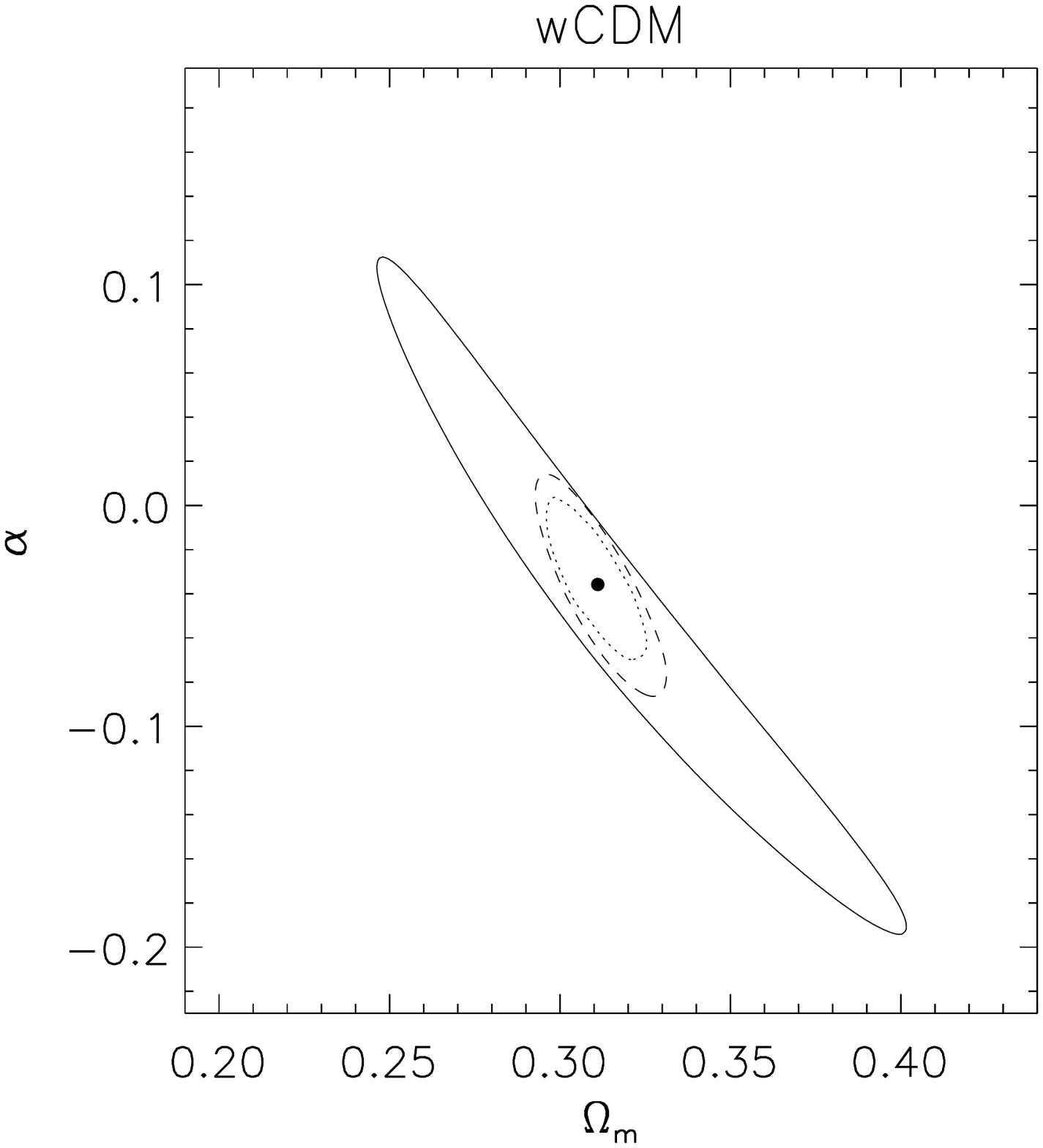}
}
\caption{Left-hand panel: 2D marginalized posterior distributions for $\Omega_m$ and $w$ in spatially flat cosmological models with a constant-$w$ dark energy EoS ($w CDM$ model). Right-hand panel: 2D marginalized posterior distributions for $\Omega_m$ and $\alpha$ (the parameter describing the evolution with $z$ of $f_{\rm gas}$, e.g. equation \ref{eqn:fgas}) in a flat $w CDM$ model. For all the panels, we show constraints arising from our X-ray-only analysis of the stacked emission measure $\widetilde{EM}(x)$ in the whole {\em Chandra} sample (solid line); and by cumulatively adding the likelihood from the Planck+WP+Union2.1 data sets (dashed line) and constraints on the evolution of $f_{\rm gas}$ from hydrodynamic simulations (dotted line) as priors. The thick long-dashed line shows the marginalized posterior distribution only from the Planck+WP+Union2.1 data sets. The contours are at the 1-$\sigma$ confidence level, while the expectation values are represented by a point and refers to the combination of all the data sets. \protect\\Notes on the complementary data sets: `{\em Planck}' refers to the CMB analysis of \protect\citet{planck2015a}; `Union2.1' refers to the SNIa compilation of \protect\cite{suzuki2012}; `WP' refers to the WMAP polarization measurements \protect\citep{bennett2013}; the constraints on the evolution of $f_{\rm gas}$ from hydrodynamic simulations refer to the results of \protect\citet{planelles2014}.}
\label{T-z3452sw}
\end{center}
\end{figure*}

The accelerated expansion of the Universe detected in the Hubble diagram for distant SNIa is one of the greatest discoveries of the past 20 years \citep{riess1998}. The acceleration can be attributed to the presence of a significant energy density component with negative pressure commonly referred as dark energy. The dark energy is $w=P/\rho$, defined as the ratio between the pressure and the energy density in the EoS of the dark energy component. A cosmological constant in the context of general relativity corresponds to a non-evolving $w = -1$. An alternative model is dynamical dark energy \citep{caldwell1998}, usually based on a scalar field, which typically has a time varying $w$ with $w \ge -1$, or alternative models which calls for modifications of general relativity on cosmological scales. Accurate measurements of $w$ in agreement with (or departing from) -1, as well as a possible evolving value of $w=w(z)$, can thus cast light on the nature of the dark energy driving the accelerated expansion of the Universe, and asses whether dark energy is truly the cosmological constant. 

In this respect, we consider spatially flat models with a constant-$w$ EoS of dark energy ($w$CDM model). This represents the most basic phenomenological extension of a $\Lambda$CDM model (where $w=-1$). Our constraint is presented in Fig. \ref{T-z3452sw} (left panel) along with independent constraints from Planck+WP+Union2.1 data sets, and the cumulative combination of priors from external data sets and hydrodynamic simulations. The different cosmological probes are in good agreement, with our constraint on the EoS being $w = -1.020 \pm 0.058$ and $\Omega_m = 0.319 \pm 0.051$ from X-ray-only data sets. We underscore again the non-negligible statistical uncertainties on the matter density parameter, which stems from the degeneracy with the hot gas fraction evolution (parameter $\alpha$, Fig. \ref{T-z3452sw} right-hand panel), as discussed in Section \ref{swefffsim3e}. For comparison, by combining only Planck+WP+Union2.1 the EoS for dark energy is constrained to $w = -1.09 \pm0.19$ (95\% confidence level), and is therefore compatible with a cosmological constant, as assumed in the base $\Lambda$CDM cosmology \citep{planck2014}. Note that the CMB alone would not strongly constrain $w$, due to the two-dimensional geometric degeneracy in these models. Hence our choice is to present directly constraints of the CMB in combination with lower redshift distance measures in order to break the degeneracy among the parameters. 

The real strength of the cluster data is, however, when they are combined with the CMB and other cosmological data sets. Indeed, adding priors on the cosmology from Planck+WP+Union2.1 data sets greatly reduces this degeneracy. We have $w = -1.010\pm 0.030$, $\Omega_m = 0.311 \pm 0.014$ and $\alpha=-0.036 \pm0.036$. While the statistical uncertainty on the matter density parameter is driven by the external data sets, we emphasize the stringent constraints on $w$ of the order of 3\%, more accurate of those from Planck+SNe or X-ray-only data, thanks to a nearly orthogonal degeneracy between ours and these external data sets (Fig. \ref{T-z3452sw}). The aforementioned degeneracy between ours and these independent data sets explains why dark energy constraints from $\widetilde{EM}(x)$ data are competitive with those e.g. from SNIa \citep{suzuki2012} and CMB \citep{planck2015a}.

Moreover, only a moderate improvement is achieved by adding priors from hydrodynamic simulations (i.e. $w = -1.009 \pm 0.026$), making our prior-free results robust towards any theoretical assumption on the $f_{\rm gas}$ evolution. Next, by incorporating different priors on $\alpha$ as priors in our Bayesian MCMC analysis [AGN, CSF and NR simulations from \cite{planelles2014}], we conservatively estimate that the bias on $w$ is $\lesssim0.3\%$.

We compare our results with \cite{vikhlinin2009c} and \cite{mantz2014}. The former, fitting cluster mass function jointly with the SNe, WMAP and BAO measurements, obtain $w = -0.991 \pm 0.045 (stat) \pm0.039 (sys)$, in excellent agreement with the present work. As for the latter work, by means of measurements of the cluster gas fraction, they obtain $w = -0.98 \pm 0.26$.

\subsection{Contraints on flat $w_z CDM$ models}\label{swevwgr3e}
In this section we consider cosmological models with an FRW metric, containing matter and dark energy, in a spatially flat Universe ($\Omega_k=0$). We adopt an evolving parametrization of the EoS of dark energy which allows $w$ to depart from -1 and to change with time. We consider here the case of a Taylor expansion of $w$ at first order in the scale factor, parametrized by $w(z)=w_0+w_az/(1+z)$ \citep{chevallier2001}. In this model, $w$ takes the value $w_0$ at the present day and $w_0+w_a$ in the high-redshift limit. This parametrization contains as special cases the cosmological constant model ($\Lambda$CDM; $w_0 =-1$ and $w_a = 0$, see Section \ref{swefffsim3e}), and constant-$w$ models ($w CDM$; $w_a = 0$, see Section \ref{swefffsim3ed}). 

\begin{figure*}
\begin{center}
\hbox{
\includegraphics[scale=0.43]{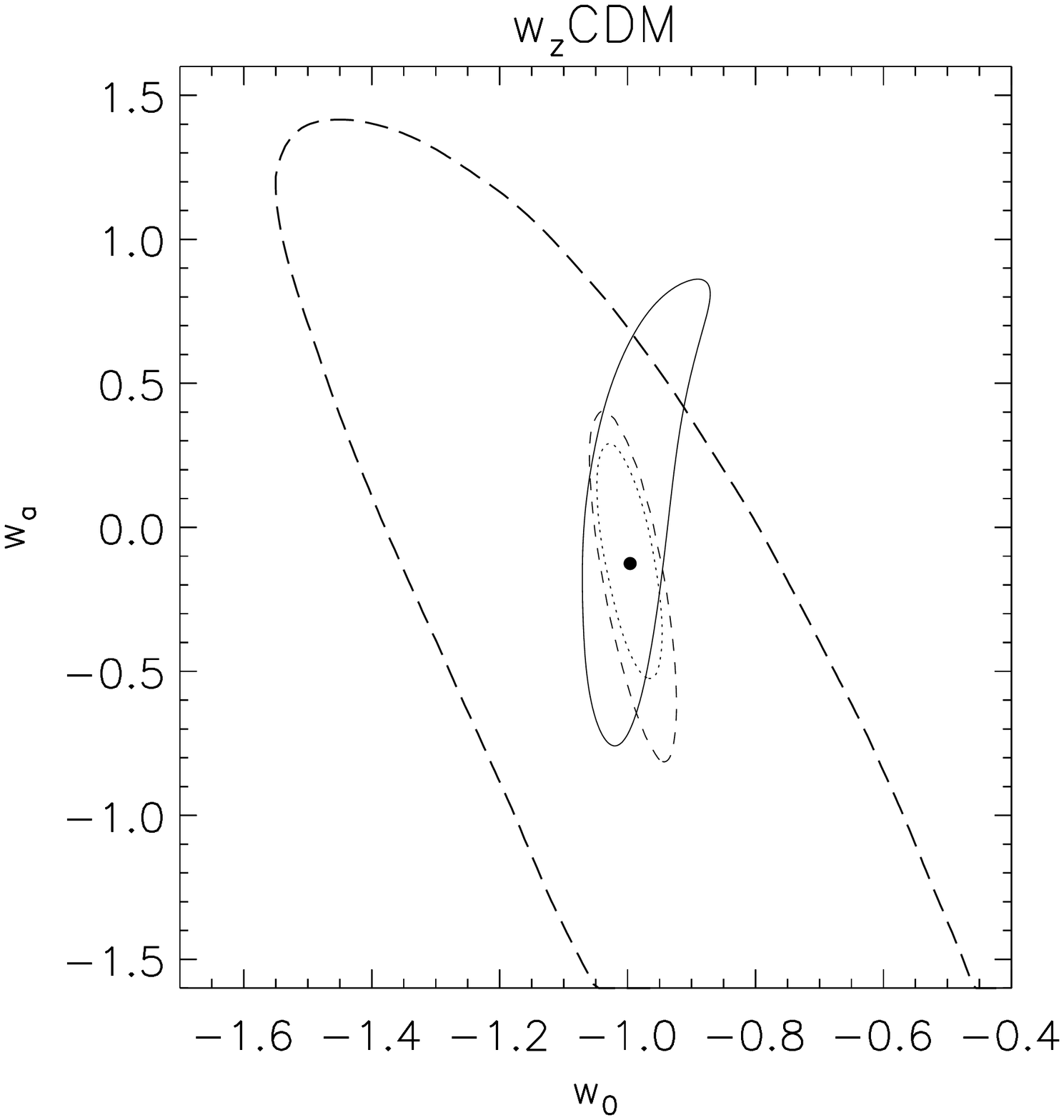}
\includegraphics[scale=0.43]{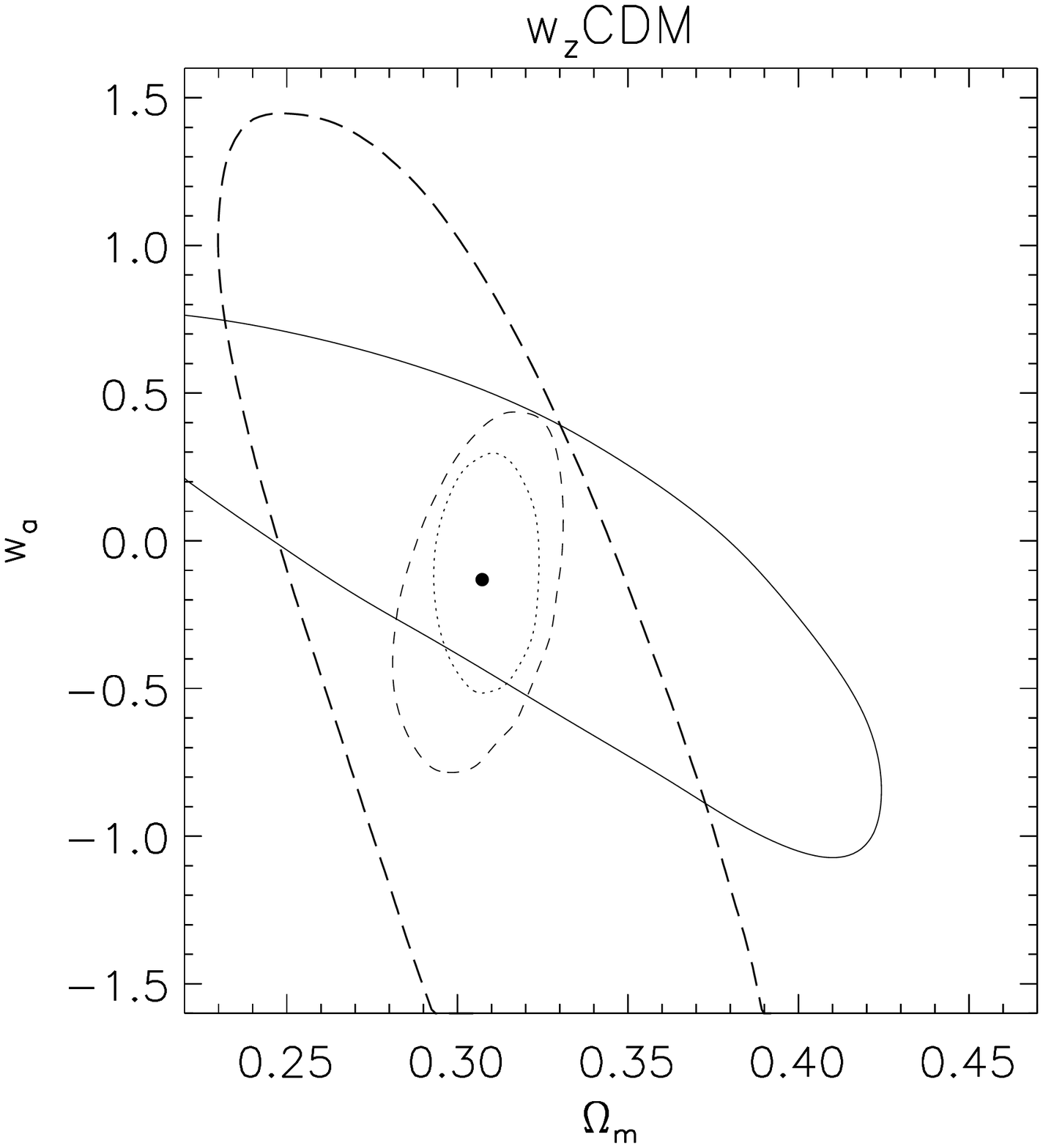}
}
\hbox{
\includegraphics[scale=0.43]{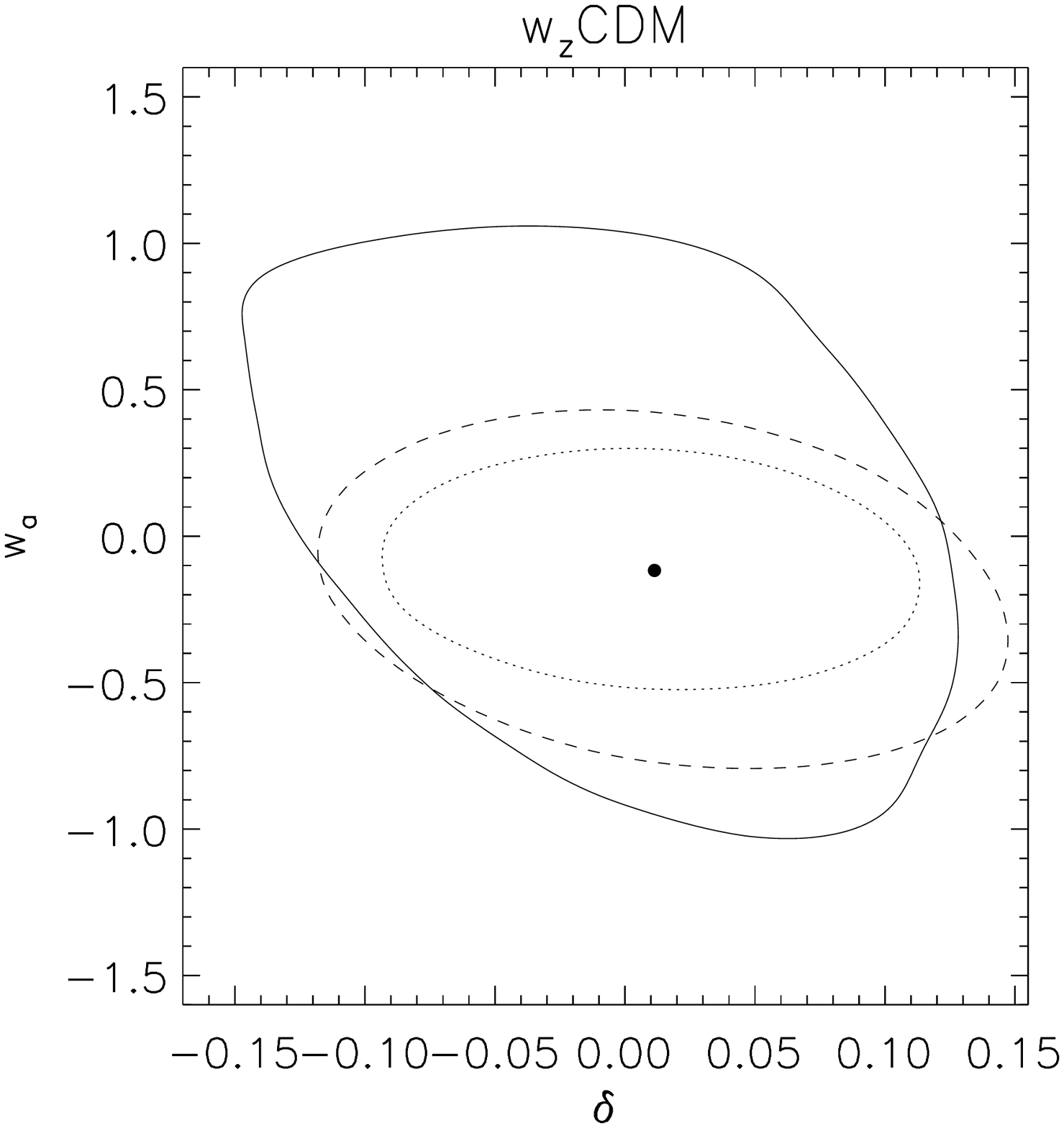}
\includegraphics[scale=0.43]{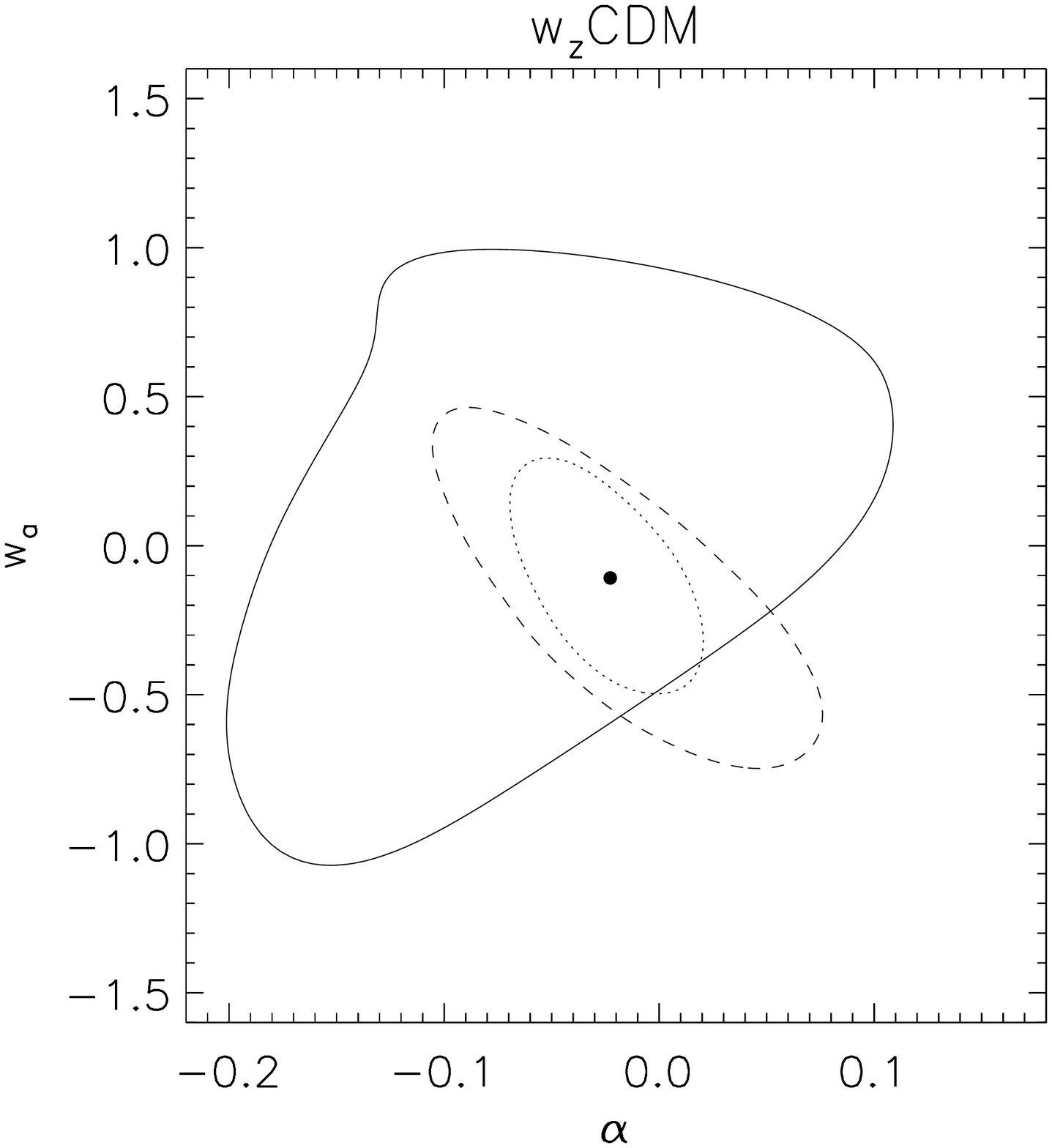}
}

\caption{Top-left panel: 2D marginalized posterior distributions for $w_0$ and $w_a$ in spatially flat cosmological models with an evolving dark energy EoS $w(z)$ (model $w_z$CDM, \protect\citealt[][]{chevallier2001}). Top-right panel: 2D marginalized posterior distributions for $\Omega_m$ and $w_a$ in a flat $w_z CDM$ model. Bottom-left panel: 2D marginalized posterior distributions for $w_a$ and $\delta$ (the parameter describing the temperature dependence of $f_{\rm gas}$, e.g. equation \ref{eqn:fgas}) in a flat $w_z CDM$ model. Bottom-right panel: 2D marginalized posterior distributions for $w_a$ and $\alpha$ (the parameter describing the evolution with $z$ of $f_{\rm gas}$, e.g. equation \ref{eqn:fgas}) in a flat $w_z CDM$ model. For all the panels, we show constraints arising from our X-ray-only analysis of the stacked emission measure $\widetilde{EM}(x)$ in the whole {\em Chandra} sample (solid line); and by cumulatively adding the likelihood from Planck+WP+Union2.1 data sets (dashed line) and constraints on the evolution of $f_{\rm gas}$ from hydrodynamic simulations (dotted line) as priors. The thick long-dashed line shows the marginalized posterior distribution only from the Planck+WP+Union2.1 data sets. The contours are at the 1-$\sigma$ confidence level, while the expectation values are represented by a point and refers to the combination of all the data sets. \protect\\Notes on the complementary data sets: ‘‘ {\em Planck}' refers to the CMB analysis of \protect\citet{planck2015a}; `Union2.1' refers to the SNIa compilation of \protect\cite{suzuki2012}; `WP' refers to the WMAP polarization measurements \protect\citep{bennett2013}; the constraints on the evolution of $f_{\rm gas}$ from hydrodynamic simulations refer to the results of \protect\citet{planelles2014}.}
\label{T-z3452}
\end{center}
\end{figure*}

Besides the dark energy EoS $w=w(z)$, the relevant cosmological parameter for the analysis of cluster data is the present-day matter density. We remember that the absolute normalization of $\widetilde{EM}(x)$ (which depends on the value of $H_0$) is a nuisance parameter in our analysis, hence we cannot constraint $H_0$.

The resulting constraints on $w_0$ and $w_a$ are shown in Fig. \ref{T-z3452} as 1-$\sigma$ confidence level contours. Curvature $\Omega_k$ is not allowed to vary, remaining fixed to zero. The data are thus consistent with the $\Lambda$CDM model ($w_0 =-1$, $w_a = 0$). Even for this generalized model, our X-ray-only analysis provides a stringent constraint $\Omega_m =0.305\pm 0.093$, $w_0=-0.995\pm 0.079$ and $w_a=-0.009\pm 0.654$.

Adding priors on the cosmology from Planck+WP+Union2.1 data sets greatly reduces the degeneracy among the fitting parameters. We have $\Omega_m=0.308\pm 0.017$, $w_0=-0.993\pm 0.046$ and $w_a=-0.123\pm 0.400$. While the statistical uncertainty on the matter density parameter is driven by the external data sets, we stress the stringent constraints on $w_a$ with an uncertainty of the order of 0.4, thanks to a nearly orthogonal degeneracy between ours and these independent data sets (Fig. \ref{T-z3452}). 

A non-negligible improvement is achieved by adding priors from hydrodynamic simulations ($\Omega_m=0.309\pm 0.010$, $w_0=-0.997\pm 0.034$ and $w_a=-0.081\pm 0.267$), with an improvement of a factor of 1.5 for the statistical uncertainties on $w_a$. However, by incorporating different constraints on $\alpha$ from hydrodynamic simulations (e.g. CSF and AGN simulations) as priors in our Bayesian MCMC analysis, we conservatively estimate that the bias on the matter density parameter and $w_0$ is $\lesssim0.4$ and $\lesssim0.3$ per cent, respectively.

We finally compare our result with \cite{mantz2014}. Fitting cluster mass function jointly with the 1-year Planck data plus WMAP polarization \citep{planck2013} they obtain $\Omega_m=0.298 \pm 0.015$,  $w_0=-1.03 \pm 0.18$ and $w_a=-0.1^{+0.6}_{-0.6}$, in excellent agreement with the present work.

\section{Constraints on the EoS of dark energy: model selection}\label{sweg6fsim}
In this section, we perform a second level of Bayesian inference, which is a model comparison. Here we wish to compare the models in the light of the data, and assign some sort of preference or ranking to our different assumptions. We will determine which cosmological model $\Lambda$CDM, $w$CDM and $w_z$CDM (Section \ref{swcdg6im3ed}) and definition of boundary radius (Section \ref{swcdfffrm3ed}) are more likely given the data. Note that this Bayesian model selection will be possible since our priors on the cosmological and astrophysical parameters are informative, defining statistical information about fitting variables.

\subsection{Comparison of $CDM$ models}\label{swcdg6im3ed}
In this section we wish to measure the relative quality of statistical cosmological models for a given set of data. Given different cosmological models for the data, we would like to determine `which model is more likely' given the data, providing the best trade-off between the goodness of fit of the model and its complexity. 

While the recent accelerated expansion of the Universe detected in the Hubble diagram for distant SNIa is one of the greatest discoveries of the past 20 years \citep{riess1998}, the question arises whether the dark energy $w=P/\rho$ (the ratio between the pressure and the energy density in the EoS of the dark energy component) is truly the cosmological constant. A cosmological constant in the context of general relativity corresponds to a constant EoS of the dark energy parameter $w = -1$. An alternative model is dynamical dark energy, or models which calls for modifications of general relativity on cosmological scales, which typically has a time varying $w$ \citep{caldwell1998}. Precise measurements of $w$ in agreement with (or departing from) -1, as well as a possible evolving value of $w=w(z)$, can thus cast light on the nature of the dark energy driving the accelerated expansion of the Universe. Only a value very close to -1 would determine that the dark energy is truly the cosmological constant. 

In this respect, we used the evidence-based criterion. Assuming that we choose to assign equal priors $P(\mathcal{H}_i)$ to the alternative models (e.g. $w_z CDM$) with respect to the reference model ($\Lambda$CDM), we rank cosmological models by evaluating the evidence ratios. Thus, the $\Lambda$CDM cosmology is more supported, showing a Bayes factor $K\sim7(13)$ with respect to the alternative model $w CDM$($w_z CDM$) \citep[for the scale for the interpretation of $K$ see][]{kass1995}. The $\Lambda$CDM model shows a Bayes factor $K\sim288$ with respect to a non-flat $\Lambda$CDM model, and thus the latter is strongly disproved.

Note that the prior on $\delta$ is non-informative, since we do not require any bound for this parameter. While the exact Bayes factor calculation may slightly depend on the allowed range of values of $\delta$, we verified that this parameter is substantially a nuisance, being consistent with zero and poorly correlated with the other parameters (see Figures \ref{T-z3452b} and \ref{T-z3452}). Thus, in our Bayes factor calculation we fixed $\delta=0$ and we combined all the data sets and priors.

Finally, we caution the reader that our analysis can constrain only a small subset of parameters (e.g. matter density and EoS of the dark energy parameter $w$) with respect to e.g. the standard spatially-flat six-parameter $\Lambda$CDM cosmology with a power-law spectrum of adiabatic scalar perturbations, or any extension of this model including e.g. dynamic dark energy, non-flat geometry etc. Therefore, our quantitative Bayes factor calculation merits some caution, although semi-quantitatively we emphasize that there is no evidence of non-constant ($w\ne -1$) or time-evolving ($w=w(z)$) dark energy given the data.

\subsection{Boundary radius definition}\label{swcdfffrm3ed}
In Section \ref{selfsim} we described the self-similar model, and we underscored that clusters are identical objects when renormalized by their mass or temperature, and rescaled according to a certain radius (e.g. $R_{200}$) with respect to the reference background density of the Universe \citep{kaiser1986}. Since cluster's mass and all the relations connecting the mass to other observable quantities depend on how one defines a cluster’s outer boundary, a robust definition of `virial' radius is clearly important. 

However, while one would like to define that boundary in order that the relationships between cluster mass and other observables are as simple as possible, clusters formed from hierarchical clustering do not have unique, or even distinct, outer boundaries. Our baseline model defines $R_{200}$ as the radius within which the mean total density is $\Delta_c=200$ times the critical density of the Universe $\rho_{c,z}$\footnote{Only in this section we will use the notation $\Delta_c$ rather than $\Delta$ to explicitly indicate an overdensity with respect to the critical density. Elsewhere we use $\Delta$ (e.g. $\Delta=200$), dropping the subscript `c', for the sake of simplicity in the notation.}. This cluster boundary definition is conventional and it is borne out of the spherical top-hat model for non-linear collapse \citep[e.g.][]{evrard2008}, since it is close to the value $18\pi^2$ for the critical Einstein-de Sitter (EdS) cosmological model (with $\Omega_m=1$). However, strictly speaking, this overdensity has also a dependence on the underlying cosmology and redshift through the parameter $\Omega_z = \Omega_{m} (1+z)^3/E_z^2$. This generalization of the parameter $\Delta_{c,z}$ can only be calculated by numerically integrating the equations of motion for the spherical top-hat collapse in a $\Lambda$CDM cosmology, with the assumption that the baryonic matter follows the DM and neglecting effects of hydrodynamics in structure formation.

\cite{eke1998} provided fitting formulae for $\Delta_{c,z}$ for $\Lambda$CDM models:
\begin{equation}
\Delta_{c,z}=  18 \pi^2 +82 (\Omega_z-1) -39 (\Omega_z-1)^2
\label{delta_z22}
\end{equation}

Thus, the relations presented in Section \ref{selfsim} can be generalized by replacing $E_z$ with the term $f_z\equiv E_{z}{(\Delta_{c,z}/\Delta)}^{1/2}$ \citep[see also][]{ettori2004}, which captures a redshift-dependent overdensity parameter $\Delta_{c,z}$. The special case of constant overdensity corresponds to $f_z\equiv E_{z}$ (see equation \ref{dewwefq3red2wws}).

In the literature $\rho_{c,z}$ has been widely used to define cluster boundaries, with a redshift-dependent or constant overdensity. In particular, $\Delta_c=500$ (fixed overdensity) has been used in the analyses of Chandra and XMM-Newton X-ray observations of galaxy clusters, since it marks the transition between the virialized region, where the X-ray signal can be measured reliably, and the outer volumes \citep{vikhlinin2006,simionescu2011,eckert2012,planck2013}. However, we can expect that the ICM in galaxy clusters shows some hydrodynamic modifications with respect to the gravity-dominated scale-free DM distribution, as well as an highly non-spherical topology of the LSS formation \citep{bohringer2012,kravtsov2012}, which are not captured by the spherical top-hat collapse model. Thus, the question arises whether the generalization of the parameter $\Delta_{c,z}$ in equation (\ref{delta_z22}) is a nuisance. For example, $N$-body simulations indicate that the dependence of the overdensity parameter on the redshift can be neglected \citep{evrard2008}. Ultimately, we wish to determine by means of observables what is the best definition of cluster boundary.

Finally, other definitions of `virial' radii hinge on the average matter density of the Universe, rather than that critical, e.g. $\rho_{m,z}=\rho_{c,z}\Omega_z$ \citep{lau2015}, which is independent of other cosmological parameters, e.g., the Hubble parameter. The fiducial radius of a cluster is thus defined as the radius of a sphere enclosing an average matter density equal to a reference overdensity $\Delta_{m}$ times an average background density $\rho_{m,z}$. Using $\rho_{m,z}$ to define DM haloes also leads to universal mass function in $N$-body simulations \citep{jenkins2001}. In this case, the relations presented in Section \ref{selfsim} can be generalized by setting $f_z\equiv (1+z)^{3/2}$ and replacing $E_{z}$ with $f_z$.

Here we are interested to compare three definitions of boundary radius, namely based on:
\begin{itemize}
 \item[1)] a constant overdensity $\Delta_{c}$ with respect to the critical density of the Universe;
 \item[2)] a redshift-dependent overdensity $\Delta_{c,z}$ with respect to the critical density of the Universe;
 \item[3)] a constant overdensity $\Delta_{m}$ with respect to the average matter density of the Universe.
\end{itemize}

In general, we can rewrite equation (\ref{dweded2}) and highlight the dependence of $\widetilde{EM}(x)$ on the cosmological parameters with these different definitions of cluster boundary:
\begin{equation}
\widetilde {EM}(x)  \propto  T_X^{-1/2} f_{\rm gas}^{-2} f_z^{-3+(1-6\beta)}\; D_{\rm a}^{1-6\beta}
\label{dewwefq3red2wws}
\end{equation}
The expansion history factor $f_z=\{ E_{z}, E_{z}{(\Delta_{c,z}/\Delta_c)}^{1/2},(1+z)^{3/2}\}$ for the aforementioned overdensities $\Delta_{c}$,$\Delta_{c,z}$ and $\Delta_{m}$, respectively, and the angular diameter distance $D_{\rm a}$ are thus the proxies of the cosmology, depending on $\Omega_m$ and $w=w(z)$. 

\begin{figure}
\begin{center}
% \epsscale{.85}
\includegraphics[scale=0.43]{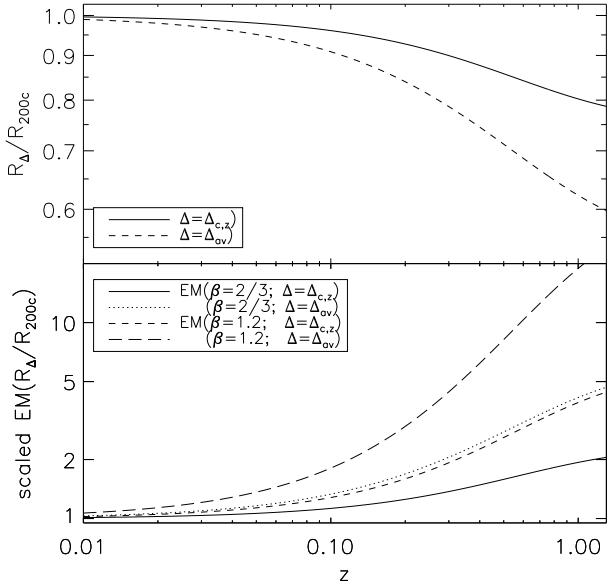}
% \plotone{ps/T-z.ps}
\caption{Upper panel: evolution of the boundary radius $R_{\Delta}$ ($\Delta=\{\Delta_{c,z},\Delta_{m}\}$) renormalized by $R_{200c}$. We arbitrarily rescaled $R_{\Delta}$ in such a way that $R_{\Delta}/R_{200c}=1$ at $z=0$ and for $\Omega_m=0.3$, following the evolution of this ratio with redshift. Lower panel: toy model (a $\beta$-model profile with slopes $\beta=2/3$ and 1.2) describing the scaled $EM$ $\widetilde{EM}(R_{\Delta}/R_{200c})$ (renormalized in such a way that it is equal to unity at $z=0$). The $EM$ profiles have been scaled according to the standard self-similar model (equation \ref{dweded2}). By rescaling the profile by $R_{200c}$, the ensuing $EM$ at the boundary radius $R_{\Delta}$ shows larger values for the high-$z$ cluster population, thus introducing unphysical departures from self-similarity.}
\label{T-zdel2}
\end{center}
\end{figure}

Here we propose to use self-similarity to discriminate which of the definition of cluster boundary is more tenable. The idea is to consider the evolution with $z$ of the physical properties, namely the scaled $EM$, of the ICM of our population of clusters by rescaling with respect to various definitions of boundary radius. By inferring $\widetilde{EM}(x)$ ($x=R/R_{\Delta}$) at different $z$, we can recover the definition of `virial' radius $R_{\Delta}$ (which enters in equation \ref{dewwefq3red2wws}) which better preserves self-similarity. In Fig. \ref{T-zdel2} we present a toy model depicting the evolution of the boundary radius $R_{\Delta}$ ($\Delta=\{\Delta_{c,z},\Delta_{m}\}$) renormalized by $R_{200c}$. We arbitrarily rescaled $R_{\Delta}$ in such a way $R_{\Delta}/R_{200c}=1$ at $z=0$, following the evolution of this ratio with redshift. As we can see, this ratio is smaller than the unity at high-$z$, that is we are probing progressively more internal regions (with respect to $R_{200c}$) as we are moving to higher redshifts. The ensuing (rescaled) $EM$ at the boundary radius $R_{\Delta}$ shows larger values for the high-$z$ cluster population, thus introducing unphysical departures from self-similarity. The effect is larger for $\Delta=\Delta_{m}$ and moving towards the outer volumes due to the steeper $EM$ profiles. In other words, if self-similarity of the X-ray observables holds by rescaling them with respect to $R_{200c}$, we should observe a break of it if we introduce other definitions of `virial' radii, with high-$z$ clusters having larger $\widetilde{EM}(x)$ with respect to the low-$z$ population. 

While we proved in our previous work \citep{morandi2015} that self-similarity is very well preserved in a framework where $\Delta=\Delta_{c}$, we caution the reader that those findings are cosmological-dependent given the assumed cosmological model. Ultimately, we would like to disentangle the evolution of the cosmic expansion history from the evolution of the baryon properties, as previously discussed (Section \ref{coerg5uded2}).

In order to determine which definition of boundary radius better preserves similarity of the profiles we calculate the Bayes factor by considering the two cases $\Delta=\Delta_{c,z}$ and $\Delta=\Delta_{c}$ in deriving $\widetilde{EM}(x)$ at various redshifts. Thus, $\Delta=\Delta_{c}$ is strongly supported by our data with respect to the alternative model, with $K\sim2\times 10^{2}$ by combining all the data sets and priors \citep[for the scale for the interpretation of $K$ see][]{kass1995}. Similar conclusions have been drawn by \cite{evrard2008,bohringer2012}. Moreover, the use of this alternative definition of boundary radius ($\Delta=\Delta_{c,z}$) would bias downwards the estimate of $\Omega_m$ ($\sim0.25$), lowering the value of the Bayesian evidence. Clearly this value of $\Omega_m$ is also in tension with the {\em Planck} cosmology.

As for $\Delta=\Delta_{m}$, \cite{lau2015} found that in the outer volumes the self-similarity of the ICM profiles is better preserved if they are normalized with respect to the mean density of the Universe, while in the inner volumes ICM profiles are more self-similar when normalized using the critical density. We point out that we cannot apply the previous Bayesian inference to this case, the models ($\Delta=\Delta_{m}$ and $\Delta=\Delta_{200c}$) being not independent and overlapping out to large radii (out to $\sim R_{500c}-R_{200c}$). A simpler approach is to repeat the exercise presented in the toy model of Fig. \ref{T-zdel2} for real data, and study the evolution of $\widetilde{EM}(R/R_{\Delta_{m}})$ versus ${\widetilde{EM}(R/R_{200c})}$ with $z$ in the outer volumes ($R\gesssim R_{200c}$). By propagating all the uncertainties related to statistical+systematic measurement errors \citep[][]{morandi2015} and to the cosmological parameters, we confirmed our previous findings that self-similarity is very well preserved with respect to a fixed critical overdensity, while it breaks if we introduce the alternative definition of boundary radius ($R_{\Delta_{m}}$). In the latter case, we observe that $\widetilde{EM}$ at high-$z$ is $\sim3-4$ times larger than the local cluster values, which is statistically disproved at the $\sim6\sigma$ statistical significance level.

In this respect, our previous results \citep{morandi2015} indicate that filaments can contribute significantly to the emission at large radii. Filaments could be more difficult to remove at high-$z$ via the method we employed \citep[see][]{morandi2015}, which may lead to higher $EM$. Subclumps could also contribute to observed biases if we do not properly remove undetected substructures, given the low signal-to-noise in the high-$z$ sources (Lau \& Nagai, personal communication). We thus repeat the analysis by masking the filaments, and we confirmed the aforementioned results but at smaller statistical significance ($\sim3\sigma$), which stems from the lower stacked signal. However, we caution the reader that we are able to constrain ICM profiles out to $R_{100c}$, thus it is still possible that beyond our boundary radius the self-similarity might better preserved if cluster observables are normalized with respect to the mean density of the Universe. 

Moreover, independent SZ analyses suggest that, for an assumed cosmology, similarity of the shapes of the pressure profiles holds out to $3\times R_{500c}\sim1.5R_{100c}$ \citep{plagge2010,planck2013}, in agreement with the present analysis.

\section{Analysis of the systematics}\label{sgnegneg53}
In this section, we tackle the impact of systematics in our analysis due to biases in the recovered physical parameters, the presence of outlier measurements and selection effects. We start by presenting a simplified toy-method to validate our cosmological analysis (Section \ref{sggg53}). We then discuss the impact of outlier measurements by discussing the results via a Bagging-MCMC analysis (Section \ref{sffbagg53}), which is an alternative method with respect to the standard MCMC analysis previously discussed. Finally, we discuss the impact of inhomogeneities of the gas distribution and cosmic filaments (Section \ref{sgdgerdg3344}).

\subsection{Validation of the cosmological analysis}\label{sggg53}
In this section we wish to determine whether our results are robust towards our stacking methodology, in particular with respect to the modelling of the background.

We validated our cosmological analysis via a simplified toy-method where we generated simulations of mock data sets including systematics. This method closely follows the methodology discussed in \citet[][e.g. Section 5.1]{morandi2015}. From the rescaled $EM$, averaged on the whole sample, we created mock azimuthally averaged source brightness images $S_{X,s}(R)$ for each observation, according to the observation aspect information, a for an assumed cosmology. We then used blank-sky fields to produce realistic backgrounds $S_{X}^{bkg}({\bf x})$ for each observation, which underwent a reduction procedure consistent with that applied to the cluster data. We thus added these backgrounds to the source brightness. In this way we can capture systematics due to both possible spatial variation of the CXB and temporal variations of the particle background spectrum. The entire procedure closely mimics the stacking analysis from which we infer the rescaled emission measure $\widetilde{EM}$ used for constraining the cosmology. We finally applied the whole cosmological analysis described in Section \ref{rjmcmc1} on the mock observations. A comparison between the true and inferred cosmological parameters reveals that the measurement errors fairly represent the error budget and our recovered parameters are not significantly biased by uncertainties in the background modelling. In particular, we conservatively estimated an upper limit to the systematics on the cosmological parameters due to background modelling $\lesssim0.5\%$.

\subsection{The impact of outlier measurements}\label{sffbagg53}
Making unbiased predictions from noisy data, possibly affected by systematics due to e.g. background modelling, is a challenging problem. We remember that in our cosmological analysis we include all the clusters, irrespective of the quality of the data, relaxation state etc. While in the previous section we proved that our results are robust towards the modelling of the background, the question arises whether the inclusion of some clusters with outlier $EM$ profiles has some ramification on the stacked emission measure $\widetilde{EM}(x;z)$ in each bin of redshift, and hence on the cosmological parameters.

For example, a cluster like A1689, characterized by deep observations ($\sim200$~ks) shows a large gas fraction (and thus larger $EM$, which is the quantity we rely on), of the order of 0.2, which is non-negligibly larger than the cosmic baryon budget. This is probably due to the presence of non-thermal pressure support and triaxiality, which boosts the X-ray surface brightness and gas fraction measurements \citep{morandi2011b,morandi2011a}. On the other hand, a cluster like A133, characterized by ultra-deep observations ($\sim2$~Ms), has a low value of the gas fraction \citep[$\lesssim 0.1$;][see, also, Vikhlinin et al., in preparation]{vikhlinin2006,morandi2014} and hence $EM$. Other clusters, e.g. the Bullet clusters, are largely unrelaxed. In this respect, our approach is different from e.g. \cite{mantz2014}, who only consider massive and morphologically relaxed systems, where the geometry, total and gas mass (and hence the {\it measured} gas fraction) can be reliably measured and reliably compared to simulations, in order to have an exquisite acknowledge of the systematics for {\it individual} measurements. Thus, we believe that the ramification of stacking such heterogeneous sample in recovering the desired cosmological and astrophysical parameters deserves further investigation. 

We emphasize that in our cosmological analysis we make use primarily of (i) the median of the distribution of the (renormalized) $EM$ profiles, which is (by definition) an outlier-resistant estimator; and (ii) we infer $\widetilde {EM}(x)$ out to large radii via stacking in bins of redshift (since we are interested in exploiting the large cosmological information `buried' in the outskirts of clusters, see Fig. \ref{T-z}) rather than using individual emission profiles. Using stacked profiles adds the benefit to infer $S_X$ out to large radii, while minimizing the impact of measurement errors for individual clusters and outliers in the stacked signal.

While our cosmological results have been mostly carried out in a standard MCMC framework (Section \ref{rjmcmc1}), here we independently run Bagging-MCMC simulations (Appendix \ref{bagging}). Bagging hinges on bootstrap with resampling, fitting the model to each sample, and then averaging the predictions to get the bagged prediction. The result is a smoother prediction that is less affected by outliers or particularly noisy data points. A Bagging-MCMC analysis can be thus used as a tool to gauge the ramification of outlier measurements (which may reflect systematics, intrinsic scatter and measurement errors in the low signal-to-noise $S_X$ in the outskirts of individual clusters) on $\widetilde {EM}(x;z)$ and hence on the cosmological parameters. We confirm very good agreement between the standard and Bagging-MCMC analysis, with systematics in the recovered cosmological parameters $\lesssim0.4\%$.

As a further test, we verified that our results are robust towards systematics by considering cluster sub-samples based on redshift, temperature, relaxation state and observing time. More in details, we repeated the cosmological analysis by considering: (i) only relatively relaxed objects with centroid shift $\le2$, this quantity being a proxy of the dynamical state of clusters; (ii) masking the cosmic filaments, as described in \cite{morandi2015}; (iii) splitting the sample in two temperature bins (i.e. greater and smaller than 7 keV) but on a coarser redshift rebinning; and (iv) dividing our sample into two subsamples, clusters with total net counts smaller and larger than the median value of the whole sample, respectively. We find very good agreement for all these cases with respect to the whole sample analysis, indicating that our baseline self-similarity assumption is tenable.

The aforementioned validation tests suggest that our proposed method is robust towards systematic due to background modelling, selection effect and the presence of outlier measurements. Compared to other techniques (e.g. $f_{\rm gas}$, mass function), the proposed method has also the major advantage that it is more direct, i.e. based on observed quantities (the $EM$) on the plane of the sky, providing an independent way (and hinging on different assumptions) with respect to the former approaches. Unlike the gas fraction or the mass function methods, no assumption of hydrostatic equilibrium and three-dimensional geometry (since we do not infer the three-dimensional gas and total mass), homogeneous gas distribution, stellar baryon fraction or model of mass function (calibrated e.g. on $N$-body simulations) are required. This stems from the fact that the absolute normalization of $\widetilde{EM}(x)$ is a nuisance parameter in our analysis, hence effect of clumpiness \citep{morandi2014}, asphericity and non-thermal pressure \citep{morandi2012a} do not have a major impact on the recovered cosmological parameters. Thus, this cosmological test is also relatively insensible to X-ray calibration systematics. We only require (weak)-similarity of the (rescaled) $EM$ profiles, which has been proved to be a tenable assumption outside the cluster core.

\subsection{The impact of inhomogeneities of the gas distribution and cosmic filaments}\label{sgdgerdg3344}
In this section we gauge the impact of inhomogeneities of the gas distribution (`clumpiness') and cosmic filaments. Observations and simulations show that the gas clumping factor $C=<\!n_e^2\!>/<\!n_e\!>^2$ lies in the range $\sim 1.3-2$ at $R_{200}$ \citep{nagai2011,morandi2013b,morandi2014}, which will bias upwards the $EM$ \citep{morandi2013b,morandi2014}. The absolute value of the clumping in each cluster-centric radius is irrelevant for our purpose, since the absolute normalization of the recovered stacked $EM$ profiles is a nuisance parameter in our analysis. However, density clumping might depend on cluster mass and/or redshift, since e.g. the contribution of cosmic filaments to clumping slightly increases with cluster mass \citep{battaglia2015}.

This could translate into biases in the recovered cosmological parameters, since we assume (weak) self-similarity of the measured $EM$ profiles, which are boosted by gas inhomogeneities ($EM\propto C<\!n_e\!>^2$). We thus repeat our cosmological analysis by: (i) correcting the measured $EM$ for the clumping factor; and (ii) masking the cosmic filaments (where most of the accretion happens, and thus they should account for most of the gas clumping) via the method we previously employed \citep[see][]{morandi2015}. For the former test, since the quality of individual observations does not allow to infer the clumping factor $C=C(R)$ \citep[e.g.][]{morandi2013b,morandi2014}, we used the predictions on $C$ from the hydrodynamic simulations of \citet[][see their equation 10 which outlines a mild mass-dependence of $C$]{battaglia2015}. By means of both these tests, we conclude that there is no significant bias in the recovered cosmological parameters due to inhomogeneities of the gas distribution and the presence of cosmic filaments ($\lesssim0.3$ and $\lesssim0.5$ per cent, respectively).

\section{Prospects from future surveys}\label{swg36fwf2z1ded}

Section \ref{swefffsim} outlines the constraints on cosmological parameters obtained from the whole cluster sample. In particular, in Section \ref{rjmcmc1} we performed a quantitative comparison of the scaled $EM$ profiles of distant clusters with a local reference profile derived from hot nearby clusters (at $z\lesssim0.2$). We observed that the amount and nature of dark energy and matter density have negligible ramifications on the scaled $EM$ at the lowest redshifts. The latter can be regarded as a baseline $EM$ nearly independent of the cosmological parameters of interest, and for which uncertainties related to the model of dark energy are thus negligible. 

In particular, the comparison of the baseline $EM$ from low-redshift cluster data with the cosmological-dependent $EM$ for high-$z$ cluster (see Fig. \ref{T-z}) produces a tight constraint on dark energy models. Here we wish to investigate the impact of high-$z$ clusters on the dark energy measurements, with the goal in mind e.g. to address the potential of future survey for constraining the EoS of dark energy.

We repeat the cosmological analysis by considering only clusters at $z<0.6$ for the $w$CDM model, and compare these results to those from our whole sample analysis. As expected, high-$z$ clusters significantly increase the constraints on the EoS of dark energy, with an improvement on $w$ by a factor of 2(1.5) in terms of statistical uncertainties with respect to the lower-$z$ ($z<0.6$) subsample, for the case where we include the external data sets (external data sets+priors on the gas fraction). 

Interestingly, when we include high-$z$ clusters and if we consider a $\Lambda$CDM model, we have only a small improvement ($\sim15\%$) in term of statistical uncertainties on the matter density parameter with respect to the lower-$z$ subsample. Clearly, the most striking ramification of including high-$z$ clusters in the cosmological analysis is to greatly improve the constraints on the EoS of dark energy, while only a slightly better accuracy is achieved on the matter density parameter. 

Future large survey should target both high-$z$ clusters and nearby objects (which provides a baseline $EM$, nearly independent of the cosmological parameters of interest) to investigate the nature of dark energy. Significant improvement in constraining the EoS of dark energy via the $\widetilde{EM}$ analysis will also necessitate the discovery of new high-$z$ ($z > 0.6$) clusters from upcoming surveys. In this respect, as we already observed (Section \ref{selfsim}), clusters of given temperature appear smaller ($R_{200}\propto E_z^{-1}$) and brighter ($EM_0\propto E_z^3$) with increasing redshift, following their self-similar evolution. This translates into a surface brightness $S_x\propto E_z^3\, (1+z)^{-4}$, which corresponds to a modest dimming at $z\sim0.6-1$ (a factor of $\sim2-3$). Therefore, only a moderate investment of observing time by X-ray telescopes will be required to observe these new high-$z$ targets, since we are interested in the stacked X-ray signal in the self-similar outskirts rather than the emission from individual sources.

A significant improvement on these constraints could be achieved, at the same time, by leveraging on priors on the gas fraction evolution from larger samples of hydrodynamic simulations. However, as we discussed, this will also require significant improvements in the physics in simulations, in order to minimize any systematic effect from the theoretical predictions.

\section{Conclusions}\label{concl22}
Measuring the EoS of dark energy is one of the largest efforts of observational cosmology. By accurately measuring $w$, we can elucidate the nature of the dark energy driving the accelerated expansion of the Universe, and assess whether dark energy is truly the cosmological constant ($w \equiv -1$).

In this work based on {\em Chandra} data, we analysed the $EM$ profiles of a sample of 320 hot ($kT > 3$ keV) galaxy clusters, covering a redshift range $z=0.056-1.24$. The derived $EM$ profiles are scaled according to the (weak) self-similar model of cluster formation, which has been used, in turn, to put constraints on the cosmological parameters (i.e. matter density and, in particular, the EoS of the dark energy parameter $w$), as well as astrophysical parameters (gas fraction evolution and its dependence upon the system temperature). The cosmological constraints are borne out of the comparison of the scaled $EM$ between nearby and higher-$z$ clusters, the correct cosmology being the one for which the various profiles at different redshifts are weakly self-similar, explicitly allowing for temperature and redshift dependence of $f_{\rm gas}$.

This cosmological test, in combination with {\em Planck}+SNIa data, allows us to put a tight constraint on the dark energy model. For a constant-$w$ model, we have $w = -1.010\pm 0.030$ and $\Omega_m = 0.311 \pm 0.014$, while for a time-evolving EoS of dark energy $w(z)$ we have $\Omega_m=0.308\pm 0.017$, $w_0=-0.993\pm 0.046$ and $w_a=-0.123\pm 0.400$. Constraints on the cosmology are further improved by adding priors on the gas fraction evolution from hydrodynamic simulations. Current data favour the cosmological constant with $w \equiv -1$, with no evidence for dynamic dark energy. 

Next, our work provided for the first time constraints on which definition of cluster boundary radius is more tenable given the data, namely when all the physical quantities are defined at a fixed overdensity with respect to the critical density of the Universe. This is especially crucial in analysis of sample of clusters spanning a wide range of redshift, where different cluster boundary definitions mostly differ (see Fig. \ref{T-zdel2}). Moreover, the most powerful constraints on current cosmological and astrophysical models of clusters (e.g. scaling relations, gas fraction, mass function and scaled $EM$) both hinge on the underlying assumption of self-similarity and arise from observations of how clusters evolve with time, which inevitably depend on how one defines their `virial' radius.

We performed a validation of our findings, providing that they are robust towards different assumptions and possible sources of systematics in our analysis, including biases due to the modelling of the background, the presence of outlier measurements, selection effects, inhomogeneities of the gas distribution and cosmic filaments. This makes this method extremely attractive, since it is more direct, i.e. based on observed quantities (the $EM$) on the plane of the sky, providing an independent way (and hinging on different assumptions) with respect to other cluster-based cosmological tests. 

The cosmological constraints derived in this analysis are in remarkable agreement with the results obtained with SNIa \citep{suzuki2012}, CMB data \citep{planck2015a}, cluster gas fraction \citep{mantz2014}, cluster growth \citep{vikhlinin2009c}, or from the combined analysis of external data sets \citep{planck2015a}, providing stringent constraints on the EoS of dark energy. This is an additional sign that we are entering an era where complementary cosmological tests provide consistent results, and we are measuring the cosmological parameters more and more accurately. It is indeed paramount that the dark energy constraints at the percent level of accuracy are obtained from combination of several independent techniques, since observational signatures of deviations in the EoS of dark energy from -1 are very small, and hence the measurements are prone to statistical and systematic errors. This not only reduces systematics but also improves statistical accuracy by breaking degeneracies in the cosmological parameter constraints.

\section*{Acknowledgements}
We are indebted to Jimmy Irwin, Matthieu Roman, Jean-Baptiste Melin, Adam Mantz, Nabila Aghanim, Maxim Markevich, Erwin Lau and Daisuke Nagai for valuable comments. A.M. gratefully acknowledges the hospitality of the Harvard-Smithsonian Center for Astrophysics and of the NASA's Goddard Space Flight Center. We acknowledge support from Chandra grants GO2-13160A, GO2-13102A, GO4-15115X and NASA grant NNX14AI29G. This work was made possible in part by a grant of high performance computing resources and technical support from the Alabama Supercomputer Authority. We thank the anonymous referee for the careful reading of the manuscript and suggestions, which improved the presentation of our work.

\begin{appendix}

\section{Bagging in MCMC}\label{bagging}

In this appendix, we synthesize the idea underlying the resampling in MCMC algorithms. The theoretical motivation is that the presence of one or more observations (in our case, the cluster $EM$) farther from the true trend can affect the fit, with severe outliers which might cause predictions on the desired parameters to become inaccurate (i.e. biased).

\citet{breiman1996} introduced {\it Bagging} (Bootstrap AGGregatING) to reduce the prediction bias (or inaccuracy) due to the presence of outliers and/or departures from Gaussianity in the measurements. Bagging can be implemented during MCMC to achieve smoother (i.e. unbiased) predictions \citep{clyde2001}. 

Bagging hinges on bootstrap with resampling, fitting the assumed model to each sample, and then averaging the predictions on the model parameters to get the bagged prediction. The bootstrap works by resampling the original data $\vect{D}=(D_1,...,D_n)$ in order to produce $M$ pseudo-data sets $(\vect{D}'_1,...,\vect{D}'_M)$, by drawing $n$ samples with equal probability and with replacement, i.e. with some elements of the original data set which may be repeated or omitted for a particular pseudo data set. In Bagging, to evaluate the best-fit model parameters, one simply fits that assumed model on each of the $M$ pseudo-data sets $\vect{D}'$, and then perform an average of the results to get the bootstrap estimate. In our case, the i$th$ data set $D_i$ is represented by the radial surface brightness $S_x(\theta)$ of the i$th$ cluster.

Going one step further, Bagging is a natural ramification of the MCMC technique, because MCMC predictions on the model parameters are themselves an average across many samples. The idea underlying Bagging is to create a Markov chain with stationary distribution equal to the posterior distribution, by re-sampling from the empirical distribution function of our measurements or observed data. We resample the original data set $\vect{D}$, with replacement, after each MCMC iteration (each full cycle through the parameters) in order to obtain $M$ pseudo-data sets $\vect{D}'$. For each of the $M$ pseudo-data sets $\vect{D}'$, the set of model parameters is thus updated sequentially via an MCMC cycle; then the full cycle is repeated in each iteration. Finally, estimates of the model parameters are taken as an average simultaneously over the resampled data $\vect{D}'$ and over the samples from the posterior distribution $P(\btheta|\vect{D}')$.

For each model ${\mathcal{H}}$ (e.g. $\Lambda$CDM, $w_z$CDM etc.), the final MCMC algorithm incorporating Bagging can be summarized as follows:
\begin{itemize}
 \item[1.] start the MCMC, with initial values and a model ${\mathcal{H}}$, on the orginal data set $\vect{D}$;
 \item[2.] cycle through all of the parameters of the model ${\mathcal{H}}$, updating the parameters themselves from conditional distributions, in order to achieve convergence to a stationary posterior distribution $P(\btheta|\vect{D},\mathcal{H})$;
 \item[3.] resample the orginal data set $\vect{D}$ by means of bootstrap, in order to obtain a new pseudo-data set $\vect{D}'$;
 \item[4.] cycle through all of the parameters of the model ${\mathcal{H}}$, updating each from conditional distributions, seeking convergence to a stationary posterior distribution $P(\btheta|\vect{D}',\mathcal{H})$;
 \item[5.] repeats steps $3-4$ until the desired number of MC samples are obtained for each model ${\mathcal{H}}$.
\end{itemize}

Clearly, this method may require a prohibitive computational time to seek convergence, since it is basically an `MCMC within an MC'. A shortcut could be to perform a fit in the aforementioned step 4. via e.g. the Levenberg-Marquardt algorithm, which is computationally significantly faster. The downside of the latter approach is that it hinges on more restrictive assumptions with respect to an MCMC framework, e.g. on the Fisher-matrix approximation. Rather, we use the standard Bagging approach by means of several independent MCMC simulations (above steps $3-4$) each performed on a different core of a large computing node. MC samples are thus obtained for MCMC simulations and then merged to infer the desired (posterior) likelihood. This required us $\sim$20,000 CPU hours to infer the cluster likelihood functions on full parameter grids for a given cosmological model.

Bagging-MCMC simulations represent a tool to gauge the ramification of outlier measurements (which may reflect systematics, intrinsic scatter and measurement errors in the low signal-to-noise $S_X$ in the outskirts of individual clusters) in the cosmological parameters (Section \ref{sffbagg53}).

\end{appendix}

% \bibliographystyle{mnras}
% \bibliography{master}

\newcommand{\noopsort}[1]{}

\end{document}